\documentclass[]{aa}
\usepackage[utf8]{inputenc}
\usepackage[T1]{fontenc}
\usepackage[dvipsnames]{xcolor}
\usepackage{txfonts}
\usepackage{amsmath}
\usepackage{placeins}
\usepackage{graphicx}
\usepackage[colorlinks=true, linkcolor=blue, filecolor=blue, citecolor=blue, urlcolor=blue]{hyperref}    
\usepackage{natbib}
\usepackage{siunitx}
\bibpunct{(}{)}{;}{a}{}{,}
\makeatletter
\renewcommand*\aa@pageof{, page \thepage{} of \pageref*{LastPage}}
\makeatother

\begin{document}
\title{Survey of Surveys. II. Stellar parameters for 23 millions of stars}
\titlerunning{SoS stellar parameters}
\author{ 
        A.~Turchi\inst{\ref{oaa}},
        E.~Pancino\inst{\ref{oaa},\ref{ssdc}},
        A.~Avdeeva\inst{\ref{oaa},\ref{inasan}},
        F.~Rossi\inst{\ref{oaa}},
        M.~Tsantaki\inst{\ref{oaa}},
        P. M.~Marrese\inst{\ref{oar},\ref{ssdc}},
        S.~Marinoni\inst{\ref{oar},\ref{ssdc}},
        N.~Sanna\inst{\ref{oaa}},
        G.~Fanari\inst{\ref{ssdc}},
        D.~Alvarez Garay\inst{\ref{oaa}},
        M. Echeveste\inst{\ref{oaa}},
        S. Nedhath\inst{\ref{oaa}, \ref{unifi}},
        S. Rani\inst{\ref{oaa}},
        E. Reggiani\inst{\ref{oaa}, \ref{unifi}},
        S. Saracino\inst{\ref{oaa}},
        L. Steinbauer\inst{\ref{oaa}, \ref{unifi}},
        G. Thomas\inst{\ref{aca}, \ref{ula}},
        F. Gran\inst{\ref{uca}},
        G. Guiglion\inst{\ref{zentrum}, \ref{max}, \ref{leibniz}}
        }
\authorrunning{A.~Turchi et al.}

\institute{INAF -- Osservatorio Astrofisico di Arcetri, Largo E. Fermi 5, 50125 Firenze, Italy\label{oaa}
\and Space Science Data Center, Via del Politecnico SNC, I-00133 Roma, Italy\label{ssdc}
\and INAF -- Osservatorio Astronomico di Roma, Via Frascati 33, 00040, Monte Porzio Catone, Roma, Italy\label{oar}
\and Institute of Astronomy, 48 Pyatnitskaya St., 119017 Moscow, Russia\label{inasan}
\and Dipartimento di Fisica e Astronomia, Università di Firenze,  Via G. Sansone 1, 50019 Sesto Fiorentino, FI, Italy\label{unifi}
\and Instituto de Astrof\'isica de Canarias, E-38205 La Laguna, Tenerife, Spain\label{aca}
\and Universidad de La Laguna, Dpto. Astrofísica, E-38206 La Laguna, Tenerife, Spain\label{ula}
\and Universit\'e C\^ote d’Azur, Observatoire de la C\^ote d’Azur, CNRS, Laboratoire Lagrange, Nice, France\label{uca}
\and Zentrum f\"ur Astronomie der Universit\"at Heidelberg, Landessternwarte, K\"onigstuhl 12, 69117 Heidelberg, Germany\label{zentrum}
\and Max Planck Institute for Astronomy, K\"onigstuhl 17, 69117, Heidelberg, Germany\label{max}
\and Leibniz-Institut f{\"u}r Astrophysik Potsdam (AIP), An der Sternwarte 16, 14482 Potsdam, Germany\label{leibniz}
}

\date{Received: \today}

\abstract
{In the current panorama of large surveys, the vast amount of data obtained with different methods, data types, formats, and stellar samples, is making an efficient use of the available information difficult.}
{The Survey of Surveys is a project to critically compile survey results in a single catalogue, facilitating the scientific use of the available information. In this second release, we present two new catalogs of stellar parameters (T$_{\rm{eff}}$, log\,$g$, and [Fe/H]).}
{To build the first catalog, SoS-Spectro, we calibrated internally and externally stellar parameters from five spectroscopic surveys (APOGEE, GALAH, Gaia-ESO, RAVE, and LAMOST). Our external calibration on the PASTEL database of high-resolution spectroscopy ensures better performances for metal-poor red giants. The second catalog, SoS-ML catalog, is obtained by using SoS-Spectro as a reference to train a multi-layer perceptron, which predicts stellar parameters based on two photometric surveys, SDSS and SkyMapper. As a novel approach, we build on previous parameters sets, from {\em Gaia} DR3 and {\em Andrae et al. (2023)}, aiming to improve their precision and accuracy.}
{We obtain a catalog of stellar parameters for around 23 millions of stars, which we make publicly available. We validate our results with several comparisons with other machine learning catalogs, stellar clusters, and astroseismic samples. We find substantial improvements in the parameters estimates compared to other Machine Learning methods in terms of precision and accuracy, especially in the metal-poor range, as shown in particular when validating our results with globular clusters. }
{We believe that there are two reasons behind our improved results at the low-metallicity end: first, our use of a reference catalog—the SoS-Spectro—which is calibrated using high-resolution spectroscopic data; and second, our choice to build on pre-existing parameter estimates from {\em Gaia} and Andrae et al., rather than attempting to obtain our predictions from survey data alone.}


\maketitle{}

\section{Introduction}
Large-scale spectroscopic surveys have significantly advanced our understanding of stellar astrophysics by providing a tremendous amount of spectra for several millions of stars, including low-medium resolution such as the RAdial Velocity Experiment \citep[RAVE;][]{rave}, the Sloan Extension for Galactic Understanding and Exploration \citep[SEGUE;][]{segue} and the Large sky Area Multi Object fiber Spectroscopic Telescope \citep[LAMOST;][]{lamost}, or at high-resolution such as the Galactic Archaeology with HERMES \citep[GALAH;][]{buder18}, the Apache Point Observatory Galactic Evolution Experiment \citep[APOGEE;][]{2024MNRAS.528.1393S} and the Gaia-ESO survey \citep{gilmore12}. The data provided by these surveys allow  us to directly derive precise estimates of key parameters such as effective temperature (T$_{\rm{eff}}$), surface gravity (log\,$g$), metallicity ([Fe/H]) and individual chemical abundances for a few millions of stars in the Milky Way. 

In recent years, these surveys spawned several works focused on Machine Learning (ML) methods (i.e. Neural Networks or simpler methods), in order to extract information from stellar spectra (\cite{cannon,payne,fabbroml,ggml,ggml2}). ML in astrophysics saw a huge development in the last decades of XX century and rose to a widespread usage in the first decades of XXI century. The term can be used as a general hat to cover different disciplines from Artificial Intelligence to Neural Networks and Computational Statistics. In general, we refer to ML techniques when speaking of algorithms that make use of heterogeneous or generally complex data to automatically ``learn'' and build a ``model'' that produces a desired output, using statistical methods.

However, high-quality spectroscopic measurements are hard to obtain on large stellar samples, requiring a lot of telescope time, and thus most data available is coming from photometric surveys, which provide observations in many different bands, that include hundreds of millions (up to billions) of stars. Few of the most important photometric surveys are the Sloan Digital Sky Survey \citep[SDSS;][]{abazajian03}, the SkyMapper Southern Sky Survey \citep[SM;][]{keller07} and the Two Micron All-Sky Survey \citep[2MASS;][]{2masssurvey}. While not as accurate as spectroscopic data, photometry can indeed be used to derive a rough estimate of T$_{\rm{eff}}$, log\,$g$ and [Fe/H], mainly by employing empirical or theoretical relations, and is used by many astronomers to analyze huge star samples. The availability of high-quality spectroscopic measurements, together with reliable estimates of distance and reddening, is of paramount importance to derive high quality estimates of the above parameters from photometric surveys.

With this in mind, we provide here the second data release of the Survey of Surveys, which presents astrophysical parameters for about 23 million stars. It contains {\em (i)} a new version of the spectroscopic Survey of Surveys \citep[SoS-Spectro, see ][]{tsantaki22} and {\em (ii)} the first version of the Survey of Suveys obtained with ML (SoS-ML). In particular, we calibrate the SoS-Spectro parameters with external, high-resolution spectroscopic parameters. For SoS-ML, we obtain stellar parameters from photometric surveys, by improving the accuracy compared to literature ML estimates, or in other words, by {\em enhancing} previous estimates.
The strength of this method is that, unlike in pure predictive approaches (without initial estimates in the input features), we do not try to produce parameter estimates starting from scratch, thus we are less impacted by large deviations of the model from the real measurement. To achieve this goal and train our ML model, we used the spectroscopic estimates of T$_{\rm eff}$, log\,$g$ and [Fe/H] provided by SoS-Spectro. We also make use of the astrometric, photometric, and spectroscopic data from {\em Gaia} DR3, to obtain parameters based on large photometric datasets such as SM for the southern hemisphere and SDSS for the northern one. 

This paper is organized as follows: in Sect.~\ref{refdatasets}, we describe our selected samples of data from {\em Gaia} DR3, SDSS, and SM; in Sect.~\ref{sec:specsos}, we calibrate the reference SoS-Spectro catalog; in Sect.~\ref{sec:nn}, we describe the algorithm used and its training; in Sect.~\ref{sec:res}, we build the SoS-ML catalog; in Sect.~\ref{sec:val}, we validate the results; and in Sect.~\ref{sec:con}, we summarize our results and draw our conclusions.

\section{Reference datasets}
\label{refdatasets}

Our analysis is based on different data sources. In the following, we describe how we selected and prepared the catalogs for each of the sources.

\subsection{Gaia}
\label{sec:gaia}

{\em Gaia} DR3 data\footnote{\url{https://gea.esac.esa.int/archive/}} \citep{gdr3} were used to facilitate an accurate cross-match between catalogs. In particular, the coordinates reported in our final catalog are in the {\em Gaia} DR3 system. We also used several indicators of stellar blending, binarity, non-stellarity, and variability for sample cleaning. Part of the information present in the {\em Gaia} catalog, such as magnitude, colors, and stellar parameters, were used as input parameters for our work. We complement the {\em Gaia} DR3 catalog with distances derived by \citet{bailer21}. Finally, we used {\tt teff\_gspphot} and {\tt logg\_gspphot} among the input parameters (Sect~\ref{sec:input}); we used the [Fe/H] estimates from \citet{andrae23} instead of {\tt mh\_gspphot} from {\em Gaia} DR3 because we noticed that they agree with the SoS-Spectro metallicities sensibly better than the {\em Gaia} ones. Specifically we observe an overall RMSE which is better by 0.15\,dex on the whole sample and 0.22\,dex better on low metallicity stars ([Fe/H]$\le$-1.5\,dex); since our goal is to estimate the SoS-Spectro measurements, this difference has a definitely positive impact.

   \begin{figure}[t]
   \centering
   \includegraphics[width=\columnwidth]{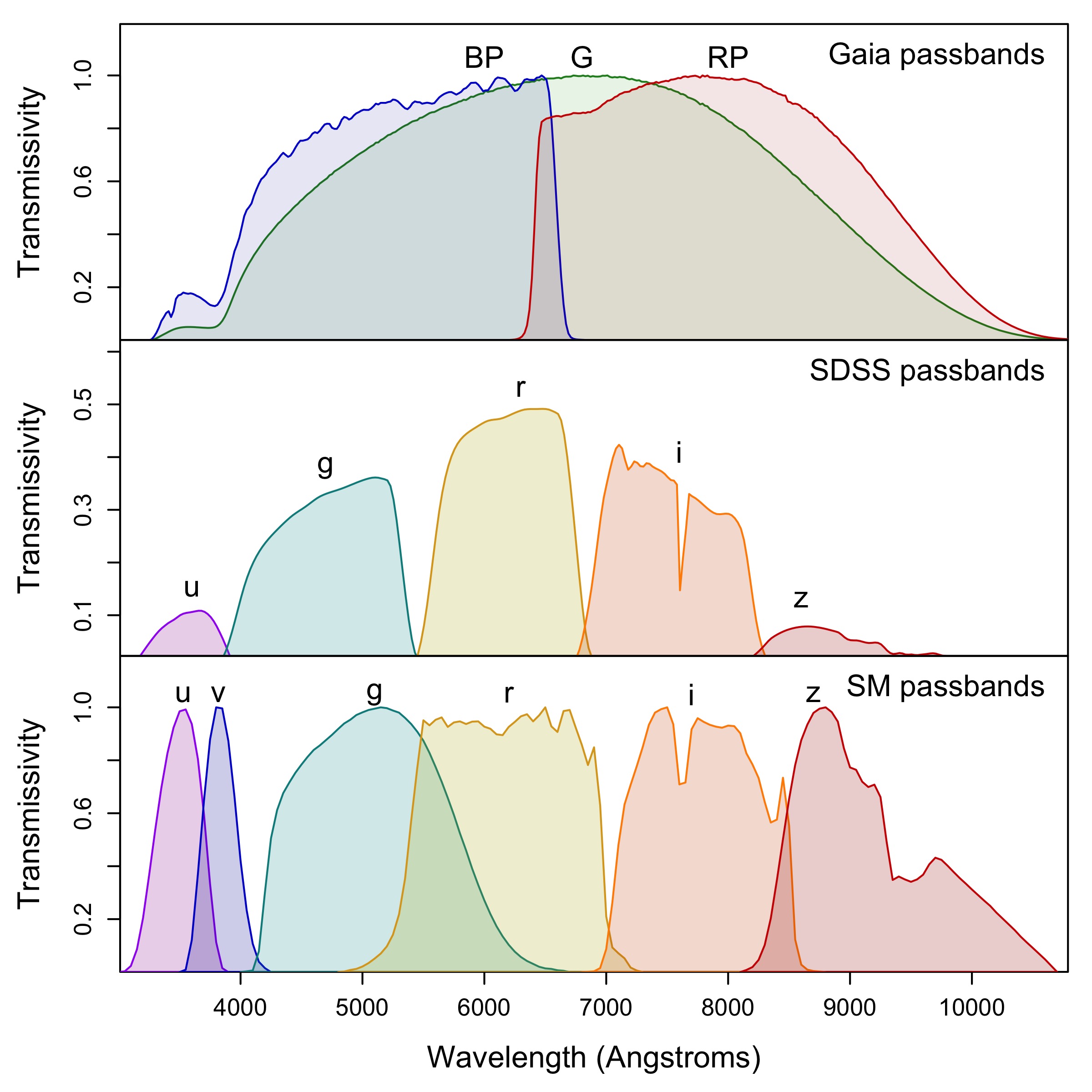}
      \caption{Photometric passbands of {\em Gaia} (top panel), SDSS (middle panel), and SkyMapper (bottom panel), obtained from the SVO filter profile service \citep[][\url{http://svo2.cab.inta-csic.es/theory/fps/}]{rodrigo24}.}
   \label{fig:band}
   \end{figure}

We applied the following selection criteria:

\begin{itemize}
    \item{sources lacking either {\tt phot\_bp\_mean\_mag} or {\tt phot\_rp\_} {\tt mean\_mag} were removed;}
    \item{sources with {\tt ipd\_frac\_multi\_peak\,$>$\,10}, {\tt ipd\_frac\_} {\tt odd\_win\,$>$\,10}, or {\tt ruwe\,$>$\,1.4} were removed, to avoid contamination by neighboring objects \citep[see also][]{lindegren21,mannucci22};}
    \item{sources with the {\tt non\_single\_star} or {\tt VARIABLE} flags were removed, as well as those with the {\tt in\_qso\_candidates} and {\tt in\_galaxy\_candidates} flags;}
    \item{we only included sources having a distance in \citet{bailer21} and in particular used the geometric distance determinations because we noted, a posteriori, that they produced slightly better results;}
    \item{we removed stars with a spectroscopic rotational broadening ({\tt vbroad}) of more than 30\,km\,s$^{-1}$, when this parameter was available, because they can have colors altered by gravity darkening and less reliable spectroscopic parameters;}
    \item{We removed all stars with a photometric temperature (teff\_gspphot) greater than 9000\,K, because they were not consistent among spectroscopic surveys.}
    \item{We removed stars with G\,$>$\,18\,mag, {\tt parallax\_error}\,$>$\,0.1 or {\tt astrometric\_sigma5d\_max}\,$>$ \,0.1, given their worse astrometric parameters. After initial tests, in fact, we noticed that $\simeq$\,13\% of the stars were clearly giants in the spectroscopic surveys, but they lied on the main sequence in the absolute and dereddened {\em Gaia} color-magnitude diagram. This conflicting information confused the algorithm, providing wrong log\,$g$ predictions for many stars in the training set. The adopted cut reduced the stars with conflicting information to about 1\% of the global sample, significantly improving our predictions.}
\end{itemize}

   \begin{figure*}[t]
   \centering
   \includegraphics[width=\textwidth]{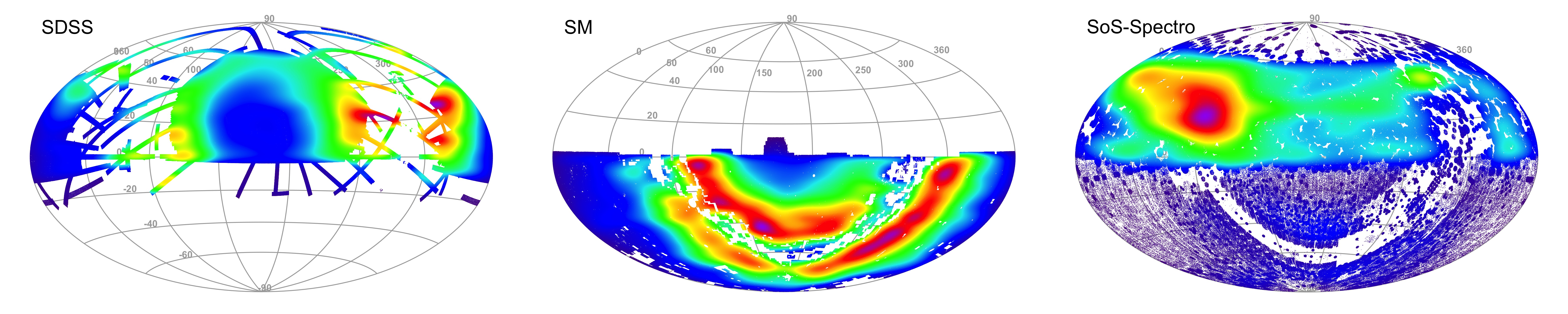}
      \caption{Sky distribution of our sample selections, after the cross-match with {\em Gaia} and all quality selections (Sect.~\ref{refdatasets}), in an Aitoff projection of RA and Dec. The color scale refers to the density of points (blue is minimum density and red maximum).}
   \label{fig:maps}
   \end{figure*}

\subsection{SDSS}
\label{sec:sdss}

For the Northern hemisphere, we used photometry from SDSS DR13\footnote{\url{https://live-sdss4org-dr13.pantheonsite.io/imaging/}} \citep{albareti17}. In DR13, no new data were added, but the previous SDSS photometric data were re-calibrated and thus considerably improved. The release contains almost half a billion objects with five-band $ugriz$ photometry (Fig.~\ref{fig:band}). We cross-matched SDSS with our {\em Gaia} DR3 selected sample (Sect.~\ref{sec:gaia}) using the software by \citet{marrese17,marrese19}. We dropped all stars in each catalog that corresponded to more than one star in the other catalog. This selection might appear very strict, but for this initial experiment, we preferred to work on the cleanest possible sample of stars. 

We removed galaxies from SDSS DR13 with {\tt type=6} and we used only well measured stars with {\tt clean=1}. To facilitate the cross-match with {\em Gaia}, we also removed stars with any of the five $ugriz$ magnitudes outside of the range 0--25\,mag. We excluded stars having less than two valid magnitude measurements and stars with reddening greater than 10~mag in the $g$ band. After these selections and after cross-matching with our cleaned sample from {\em Gaia} DR3, our SDSS dataset contains around 9 millions of stars, distributed in the sky as shown in Fig.~\ref{fig:maps}.

We computed the absolute magnitudes using {\em Gaia} distances $D$ \citep{bailer21} and the extinction coefficients $A_{B}$ from the SDSS catalogue in each band $B$ using the equation:
\begin{equation}
\label{eq:absmag}
    M_{B}= m_{B} - 5~\log_{10}{D} +5 - A_{B}
\end{equation}
where $M_{B}$ and $m_{B}$ are the absolute and relative magnitudes in each band $B$, respectively. We used the absorption coefficients $A_{B}$, provided in the SDSS catalog for each band. 

\subsection{Skymapper}
\label{sec:sm}

For the southern hemisphere, we rely on SkyMapper DR2\footnote{\url{https://skymapper.anu.edu.au/table-browser/dr2/}} \citep{huang21}, which contains photometry in the six {\em uvgriz} bands (Fig.~\ref{fig:band}) for half a billion stars. We cross-matched SM with our selection of sources from {\em Gaia} DR3 using the software by \citet{marrese17,marrese19} and similarly to the case of the SDSS, we removed from each catalog all stars that corresponded to more than one star in the other catalog. We computed absolute magnitudes as in the case of SDSS (Eq.~\ref{eq:absmag}). We used E(B--V) provided in the SM catalog, which was obtained from the maps by \citet{schlegel98}, and we obtained the $A_B$ absorption in each band $B$ following the SM recipes\footnote{\url{https://skymapper.anu.edu.au/filter-transformations/}}. We removed stars with reddening greater than 10~mag in the $g$ band.
To remove most non-stellar objects, we use the criterion {\tt classStar\,>\,0.8}. We also removed problematic measurements with the conditions {\tt flag=0} and {\tt flagPSF=0}. Similarly to the SDSS case, we removed stars outside of the range 0--25\,mag. After these selections and after cross-matching with our cleaned {\em Gaia} DR3 sample, we remained with around 10 millions of stars, distributed in the sky as shown in Fig.~\ref{fig:maps}. 

   \begin{figure*}[t]
   \centering
   \includegraphics[width=1.\textwidth]{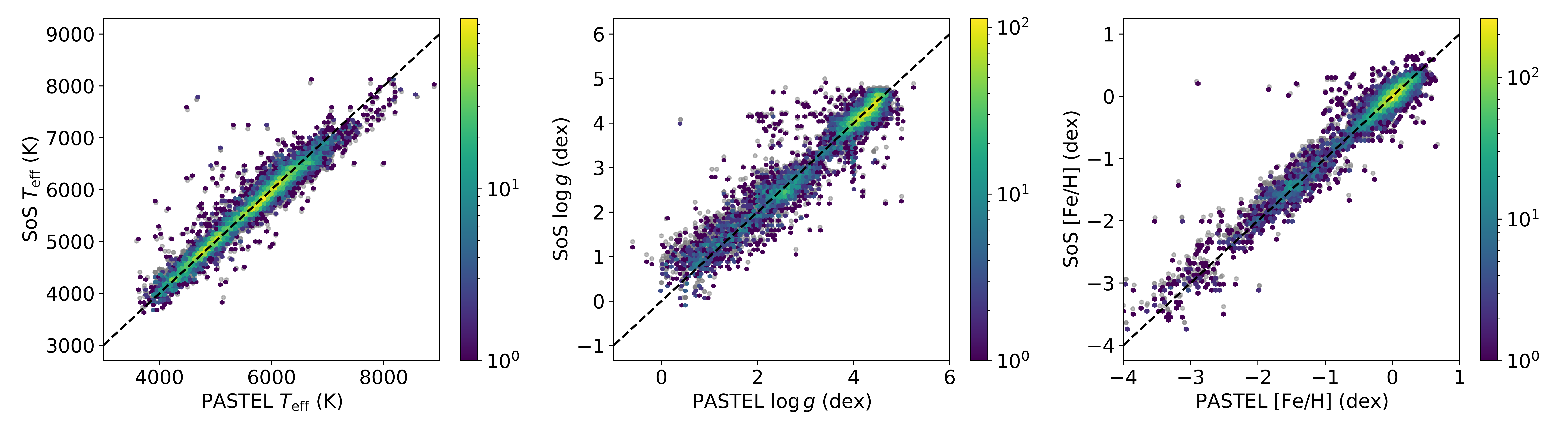}
      \caption{Comparison of the SoS-Spectro atmospheric parameters \citep[][after calibration]{tsantaki22} with the literature compilation of high-resolution studies in the PASTEL database \citep{soubiran16}, for $\simeq$15\,000 measurements, corresponding to $\simeq$14\,000 unique stars in common. See Sect.~\ref{sec:specsos} for details. Lighter color corresponds to higher data density per pixel. Grey points in the background represents the uncorrected data before applying Eq.~\ref{eq:fitpastel}.}
   \label{fig:pastel1}
   \end{figure*}

\section{Spectroscopic Survey of Surveys}
\label{sec:specsos}

Our sample of stars with known spectroscopic T$_{\rm{eff}}$, log\,$g$, and [Fe/H] (which we will use as ``labels'', according to the ML conventions) was obtained from the first SoS data release \citep{tsantaki22}\footnote{\url{http://gaiaportal.ssdc.asi.it/SoS/query/form}}. Although the release only contains radial velocities (RV), a preliminary catalog of stellar parameters was also prepared (hereafter SoS-Spectro) and used to study the trend of RV in each survey against different parameters. This was built using the following spectroscopic surveys: APOGEE DR16 \citep{ahumada20}; GALAH DR2 \citep{buder18}; Gaia-ESO DR3 \citep{gilmore12}; RAVE DR6 \citep{steinmetz20b,steinmetz20a}; and LAMOST DR5 \citep{deng12}. The homogenization procedure was very simple \citep[see equations in Appendix~A by][]{tsantaki22}: the T$_{\rm{eff}}$ of LAMOST and the [Fe/H] of RAVE were calibrated and then an average of all surveys was taken. The typical uncertainties of SoS-Spectro are nevertheless quite small, of the order of $\lesssim$100\,K in T$_{\rm{eff}}$ and of $\simeq$0.1\,dex in  log\,$g$ and [Fe/H] and the catalog contains almost 5 millions of stars.

SoS-Spectro was already successfully used to characterize Landolt and Stetson secondary standard stars \citep{pancino22}, with excellent results even for difficult parameters such as low log\,$g$ or [Fe/H]. The final SoS-Spectro parameters,  homogenized and recalibrated as described below, are made available along with the ones derived from the SDSS and SM photometry using ML (see Sect.~\ref{sec:res}). The few duplicates that were present in the original SoS preliminary catalog were removed and are not considered in this study, nor in the final catalog.

\subsection{SoS-Spectro external calibration}
\label{SoS-extcalin}

Before proceeding, we compared the SoS-Spectro parameters with a high-resolution compilation from the literature, the PASTEL database \citep{soubiran16}. The resulting median and median absolute deviations (hereafter MAD\footnote{Defined as $\mathrm{MAD} = k ~ \mathrm{median}(|e-\mathrm{median}(e)|)$, where $e$ is the vector of all the absolute values of the errors $e=\rm{abs}(x_{\rm{pred}}-x_{\rm{true}})$, and $k$ is a scaling factor which is added to rescale the MAD to the standard deviation (i.e., $k\simeq 1.4826$) for normally distributed data.}), in the sense of SoS-Spectro minus PASTEL, are: $\Delta$T$_{\rm{eff}}$\,=\,$-$9\,$\pm$\,143\,K, $\Delta$log\,$g$\,=\,$-$0.05\,$\pm$\,0.21\,dex, and $\Delta$[Fe/H]\,=\,0\,$\pm$\,0.11\,dex, which are quite satisfactory. However, a large spread and overestimation are present in the log\,$g$ comparison for giants (up to 2\,dex), which appears to worsen at lower metallicity. We have verified that the same behavior appears in the individual surveys when compared with PASTEL \citep[see also Figs.~3 to 8 by ][]{soubiran22}, and therefore, it is not a result of our re-calibration and merging procedure, rather it is a consistent and systematic difference between PASTEL and each of the considered spectroscopic surveys. We suspect that the problem could be related to the practical difficulties of measuring correct abundances for metal-poor giants, but based on this comparison it is difficult to decide whether the problem lies in PASTEL or in the surveys. We also observe that above $\simeq$7000\,K the T$_{\rm{eff}}$ agreement between PASTEL and SoS-Spectro breaks down. This is due to the general disagreement of the spectroscopic surveys from each other in this temperature regime. Therefore, we expect the agreement to improve once we apply a more sophisticated homogenization and external re-calibration procedure in future releases of the SoS-Spectro catalog.

We also compared our [Fe/H] results with data from open and globular clusters (see also Sect.~\ref{clustervalid}) and observed the same trend as in the PASTEL comparison. Based on this, we decided to calibrate our SoS-Spectro results against PASTEL using a three-parameter linear fit. We fitted the difference SoS-Spectro minus PASTEL with the following function:
\begin{equation}
\label{eq:fitpastel}
    \Delta f =a + b\cdot \rm{T}_{\rm{eff}}^{SoS} + c\cdot \log g^{SoS} + d\cdot\rm{[Fe/H]}^{SoS}
\end{equation}

This fit was computed for $\simeq$15\,000 measurements, corresponding to $\simeq$14\,000 unique stars in common.
In Fig.~\ref{fig:pastel1}, we show the result of the above correction.
On average, the computed error correction values are -36.8~K for T$_{\rm{eff}}$, 0.05~dex for log\,$g$, and -0.008~dex for [Fe/H]. The fitted correction coefficients for T$_{\rm{eff}}$, log\,$g$, and [Fe/H] are summarized in Table \ref{tab:fit_coefficients}. The deviations at low log\,$g$ and [Fe/H] were more than halved after the re-calibration on PASTEL. 

\begin{table}
    \centering
    \caption{Fitted correction coefficients for T$_{\rm{eff}}$, log\,$g$, and [Fe/H].}
    \label{tab:fit_coefficients}
    \begin{tabular}{lcccc}
        \hline
        \hline
        Parameter & a & b & c & d \\
        \hline
        T$_{\rm{eff}}$ & 112 & -6.66 × 10$^{-3}$ & -28.1 & -40.7 \\
        log\,$g$ & 0.159 & -2.21 × 10$^{-5}$ & -0.00242 & -0.0689 \\
         Fe/H & 0.0769 & -3.84 × 10$^{-7}$ & -0.0186 & -0.0262 \\
        \hline
    \end{tabular}
\end{table}

\subsection{Systematics due to survey releases in SoS-Spectro}
\label{sec:wiggles}

It is important to note that the SoS-Spectro parameters presented in this study are not based on the latest releases of spectroscopic surveys and {\em Gaia}. The latest releases contain several improvements in the data analysis pipelines, which are expected to yield more accurate spectroscopic parameters. Preliminary tests revealed some biases in the SoS-Spectro parameters, particularly when plotted against the absolute G-band magnitude. These biases are attributed to differences in the data releases of individual surveys (see Appendix~\ref{sec:allmaps}), with observed oscillations as a function of absolute magnitude in the following ranges: about 10--150~K in T$_{\rm{eff}}$, 0.1--0.3~dex in log\,$g$, and 0.02--0.15~dex in [Fe/H] (see Fig.~\ref{fig:sos_bias_galah_full}).

\section{Neural network model}
\label{sec:nn}

In our previous work \citep{pancino22}, we applied a few simple ML algorithms (i.e. Random Forest, K-Neighbours, Support Vector Regression, or SVR) to a sample of around 6000 secondary standard stars from Landolt and Stetson photometry, with excellent results. However, in this study the sample size is larger by a few orders of magnitude and we immediately realized that even the simplest MLP (multi-layer perceptron) network quickly outperformed other simpler ML techniques such as SVR.

MLP is a non-linear technique that uses artificial neurons (thus often called Artificial Neural Network, ANN), each with an activation function that depends on an input. A matrix of neurons defines a layer, and multiple layers can be stacked (the output of each layer serving as the input to the next one). Each neuron is connected with all the inputs and specific weights determine which and how an input contributes to activate the neuron itself to produce an output, based on a non-linear activation function. Training is performed on a dataset sample in order to converge the weights to the optimal numbers that reduce as much as possible the error on the output, using what is generically called a loss function (i.e. mean error, RMSE, and so on).
MLP usually performs better the larger the training dataset is, but is less interpretable with respect to simpler methods.

\subsection{Network design}

We used the Keras python interface \citep{keras} for the TensorFlow library \citep{tensorflow}. The MLP network is built by sequentially stacking a fully connected (dense) layer with a Leaky Rectified Linear Unit activation function (Leaky ReLU), a batch normalization layer, and a dropout layer\footnote{The advantage of this choice is that the whole training procedure was run on a simple desktop PC with consumer hardware. After careful optimization, each training run took only a few hours of CPU time.}. The batch normalization layer maintains the mean output close to zero with a standard deviation equal to one. The Leaky ReLU was chosen since it allows for a better performance of the network and a better convergence. The Leaky ReLU together with normalization helps compensating the vanishing gradient problem, where some coefficient gets smaller and smaller the deeper the network and eventually impairs the ability to efficiently use all layers to learn from input. Both batch normalization and dropout layers help to avoid overfitting (i.e., a poorly generalized solution).

We selected a total of 18 hidden layers of varying size between 80 and 160 elements with the smaller ones at the extremes and the larger ones in the middle (i.e. diamond shape). During training, we varied the loss function by allowing the model to minimize the mean absolute error for the first 1/5 of training time. During the rest of the 4/5 of the training time the model was instead minimizing the symmetric mean absolute percentage error (SMAPE), which is defined as:
\begin{equation}
    f(y,y_{\mathrm{pred}})=\frac{2~\lvert y-y_{\mathrm{pred}} \rvert}{\lvert y \rvert + \lvert y_{\mathrm{pred}} \rvert + \epsilon}
\end{equation}
where $y$ is the true (reference) value, $y_{pred}$ is the value predicted by the network, and $\epsilon=10^{-3}$ is a small value introduced in order to avoid explosion of the function.
This choice allowed optimizing convergence by first reducing the maximum error over the whole dataset and trying to minimize the effect of outliers, and finally to reduce the relative error in a second phase.

\begin{table}[t!]
\caption{Features used for Gaia-SDSS dataset (see Sect.~\ref{sec:input}).}
\label{tab:gsd_in}
\centering                         
\begin{tabular}{ll}        
\hline\hline                
Parameter            & Description \\
\hline  
r\_med\_geo          & Distance \citep{bailer21} \\
phot\_bp\_mean\_mag  & {\em Gaia} DR3 BP magnitude \\
phot\_rp\_mean\_mag  & {\em Gaia} DR3 RP magnitude \\
uabs                 & Absolute and dereddened $u$ magnitude \\
uabs\_nored          & Absolute $u$ magnitude \\
gabs                 & Absolute and dereddened $g$ magnitude \\
gabs\_nored          & Absolute $g$ magnitude \\
rabs                 & Absolute and dereddened $r$ magnitude \\
rabs\_nored          & Absolute $r$ magnitude \\
iabs                 & Absolute and dereddened $i$ magnitude \\
iabs\_nored          & Absolute $i$ magnitude \\
zabs                 & Absolute and dereddened $z$ magnitude \\
zabs\_nored          & Absolute $z$ magnitude \\
psfMagErr\_u         & PSF magnitude error in $u$ \\
psfMagErr\_g         & PSF magnitude error in $g$ \\
psfMagErr\_r         & PSF magnitude error in $r$ \\
psfMagErr\_i         & PSF magnitude error in $i$ \\
psfMagErr\_z         & PSF magnitude error in $z$ \\
teff\_gspphot        & {\em Gaia} DR3 {\tt teff\_gspphot} \\
logg\_gspphot        & {\em Gaia} DR3 {\tt logg\_gspphot} \\
mh\_andrae           & [Fe/H] from \citet{andrae23} \\
teff\_gspphot\_err   & {\em Gaia} DR3 {\tt teff\_gspphot} error \\
logg\_gspphot\_err   & {\em Gaia} DR3 {\tt logg\_gspphot} error \\
mh\_andrae\_err      & [Fe/H] error \citep{andrae23} \\
\hline 
\end{tabular}
\end{table}

\begin{table}[t!]
\caption{Features used for Gaia-SM dataset (see Sect.~\ref{sec:input}).}
\label{tab:gsk_in}
\centering                         
\begin{tabular}{ll}        
\hline\hline                
Parameter   &   Description\\
\hline  
r\_med\_geo          & Distance \citep{bailer21} \\
phot\_bp\_mean\_mag  & {\em Gaia} DR3 BP magnitude \\
phot\_rp\_mean\_mag  & {\em Gaia} DR3 RP magnitude \\
uabs                 & Absolute and dereddened $u$ magnitude \\
uabs\_nored          & Absolute $u$ magnitude \\
vabs                 & Absolute and dereddened $v$ magnitude \\
vabs\_nored          & Absolute $v$ magnitude \\
gabs                 & Absolute and dereddened $g$ magnitude \\
gabs\_nored          & Absolute $g$ magnitude \\
rabs                 & Absolute and dereddened $r$ magnitude \\
rabs\_nored          & Absolute $r$ magnitude \\
iabs                 & Absolute and dereddened $i$ magnitude \\
iabs\_nored          & Absolute $i$ magnitude \\
zabs                 & Absolute and dereddened $z$ magnitude \\
zabs\_nored          & Absolute $z$ magnitude \\
e\_u\_psf            & PSF magnitude error in $u$ \\
e\_v\_psf            & PSF magnitude error in $v$ \\
e\_g\_psf            & PSF magnitude error in $g$ \\
e\_r\_psf            & PSF magnitude error in $r$ \\
e\_i\_psf            & PSF magnitude error in $i$ \\
e\_z\_psf            & PSF magnitude error in $z$ \\
teff\_gspphot        & {\em Gaia} DR3 {\tt teff\_gspphot} \\
logg\_gspphot        & {\em Gaia} DR3 {\tt logg\_gspphot} \\
mh\_andrae           & [Fe/H] from \citet{andrae23} \\
teff\_gspphot\_err   & {\em Gaia} DR3 {\tt teff\_gspphot} error \\
logg\_gspphot\_err   & {\em Gaia} DR3 {\tt logg\_gspphot} error \\
mh\_andrae\_err      & [Fe/H] error \citep{andrae23} \\
\hline 
\end{tabular}
\end{table}

For each required output parameter (T$_{\rm{eff}}$, log\,$g$, and [Fe/H]) we trained a different network, due to performance reasons on limited hardware, so the output layer is of course only one element wide. In the future, we will experiment with more complex methods also exploiting the interdependence of the three parameters, but the current results are satisfactory (see also Sect.~\ref{sec:val}).

Other precautions were taken in order to allow the network to provide a robust response even when some of the input parameters are missing. Specifically, we added a flag for each parameter (thus doubling the input size) that takes a boolean value to indicate the presence of a valid measure for the  corresponding parameter. Then we proceeded to substitute all missing parameters with zeros, which allowed the network to operate even in the presence of missing values. Since the network knows if a measure is available or not, it can take this information into account when processing the input and produce an output even in presence of missing data, which is a common occurrence in large and heterogeneous catalogues.

\subsection{Choice of input parameters (``features'')}
\label{sec:input}

The final choice of the input parameters (i.e. ``features'', in the ML lexicon) was made following two different strategies. On the one hand, we tried a ``brute-force'' approach by adding as many input parameters as possible and discarding only the ones that were clearly irrelevant (such as source IDs and similar).
On the other side, we selected a list of parameters that are commonly considered relevant, from the astrophysical point of view, to predict the quantities of interest (for example, dereddened colors, i.e., multiband magnitudes, for temperature prediction).
In the end, we followed an intermediate approach by adding to the physically relevant list of parameters the ones that, in the brute-force list, appeared to improve the result by at least a percent of the error. We also removed some of the apparently relevant parameters that appeared to provide a negligible contribution to the final accuracy, to reduce computing times.

In tables \ref{tab:gsd_in} and \ref{tab:gsk_in}, we report the selected input parameter lists for the SDSS and SM samples, respectively, which are slightly different mainly due to the addition of the $v$ band in SM. For each band, we included both the reddened and dereddened absolute magnitudes.
This is because we noticed that the two most important parameters for our algorithm were distance and reddening. Since reddening can sometimes be very high (we got values up to 500 in $g$ band for a few stars, before filtering), we observed a small but consistent increase in performance (around 10\%) with the addition of the dereddened magnitudes, perhaps because the algorithm can take into account large reddening deviations and compute the prediction accordingly.
Of course, due to the the parameter selection in Sec. \ref{refdatasets}, the applicability of the ML method discussed in this work is limited by the same constraints. This means that we apply the method only in the presence of input parameters that satisfy a certain quality standard.

   \begin{figure*}[t!]
   \centering
   \includegraphics[width=\textwidth]{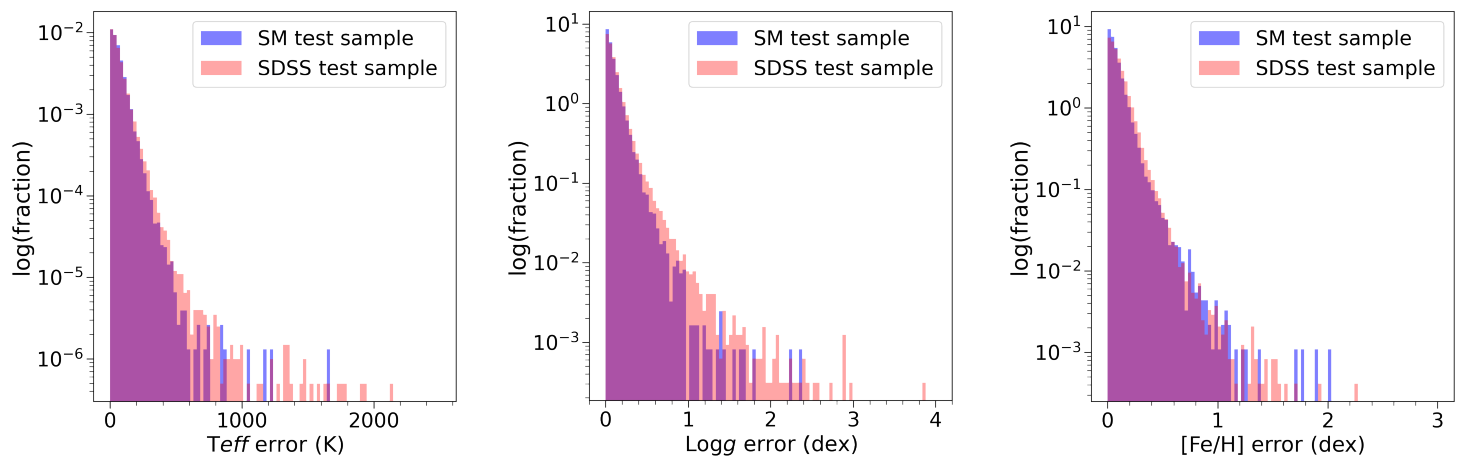}
       \caption{Distribution of ML prediction errors on the Test dataset in the SM sample case (blue) and the SDSS sample case (red). The y scale indicates the logarithm of the fraction of the Trainig dataset in each bin (total of 100 bins).}
   \label{fig:disterror}
   \end{figure*}

Given the typical problems encountered in massive ML predictions of astrophysical parameters and in spectroscopic surveys, which normally concern metal-poor stars and red giants, or hot stars, we decided to test a different approach. We thus also added estimates of T$_{\rm{eff}}$, log\,$g$, from {\em Gaia} DR3, and [Fe/H] from \citet[][see also Sect.~\ref{sec:gaia}]{andrae23}, to the list of input parameters, 
since our preliminary tests showed that the algorithm would benefit from having an initial estimate. As we will show in section \ref{sec:val}, this also allowed us to improve the predictions compared to that initial estimate.

Additionally, we noticed that adding the error estimates for the same parameters consistently increases the performance by a few percent, probably because errors helped the algorithm in discriminating how much trust to place in each provided estimate.

\begin{table}
    \centering
    \caption{Errors of the ML-predicted variables on the test sample, with respect to the SoS spectroscopic ones.}
    \label{tab:statsres}
    \begin{tabular}{lccc}
    \hline
    \hline
        & T$_{\rm{eff}}$ & log\,$g$ &  [Fe/H] \\
        & (K) & (dex) & (dex) \\
    \hline
    \multicolumn{4}{c}{{\em SDSS test sample:}} \\
        Mean & 69 & 0.11 & 0.10 \\
        Median & 50 & 0.08 & 0.07 \\
        $\sigma$ & 73 & 0.15 & 0.10 \\
        MAD & 47 & 0.07 & 0.07 \\
    \hline
    \multicolumn{4}{c}{{\em SM test sample:}} \\
        Mean & 63 & 0.10 & 0.09 \\
        Median & 48 & 0.06 & 0.06 \\
        $\sigma$ & 62 & 0.11 & 0.09 \\
        MAD & 44 & 0.06 & 0.06 \\
        \hline
    \end{tabular}
\end{table}

\subsection{Training and Testing}
\label{traintest}

We crossmatched our previously selected SDSS and SM samples with SoS-Spectro, obtaining two different samples of around 300\,000 and 800\,000 stars, respectively. These samples provide us the spectroscopic measurements for T$_{\rm{eff}}$, log\,$g$, and [Fe/H] that we will use as ``labels'' to train the model. On our first attempts, we observed that our model performed poorly (with an evident overestimation) on very low metallicities ([Fe/H]\,$\lesssim$\,--2\,dex), because both training samples had very few stars in that metallicity range. This is a common problem for spectroscopic surveys and ML methods (see Sect.~\ref{app:vmp} and Fig.~\ref{fig:metal_sum}). We thus enriched the SoS-Spectro reference sample by adding metal-poor stars from the PASTEL catalog, given the good agreement with the recalibrated SoS-Spectro. We selected stars with [Fe/H]\,$<$\,--1\,dex in common with the SDSS and SM samples, yielding an additional 139 and 267 stars, respectively. While the numbers may seem small, they increased by an order of magnitude the population of very metal-poor stars. As detailed in Appendix~\ref{app:vmp}, the addition of these very metal-poor stars did not solve the problem faced by spectroscopic surveys and ML methods in this metallicity range, but it did improve the [Fe/H] performance over the entire metallicity range (Fig.~\ref{fig:metal_sum}).

On the previously defined SDSS and SM training samples, where we have the SoS (and very few PASTEL) measurements for T$_{\rm{eff}}$, log\,$g$, and [Fe/H] to act as a label, we trained two different networks reserving 80\% of the sample for the training phase, 10\% for the internal validation phase, and 10\% for the final testing phase.
We monitored that the performance on the testing sample was never too far from the performance obtained on the training and validation samples in order to avoid overfitting.
The algorithm iterates over the training dataset a certain number of times (each iteration called epoch) and tests the value of the loss function on the validation subset (which is not used for the training itself). In the training phase, the system has access to the desired output measurements (i.e. ``labels'', the SoS-Spectro T$_{\rm{eff}}$, log\,$g$, and [Fe/H]) and will try to modify the coefficients associated to each layer in order to obtain the desired values as an output, minimizing the loss function on the validation sample. After a sufficient level of performance is obtained, the final network performance is tested on the test subset, which is completely independent from the previously used data.

In principle the reference dataset, SoS-Spectro, is the "true measure" that we want to reproduce. However, it is clear (Sects.~\ref{SoS-extcalin} and \ref{sec:val}) that the reference catalog itself has its own random deviations and biases, which will be faithfully reproduced by the MLP. In other words, the algorithm cannot perform better than the reference dataset. Therefore, any error estimate on the MLP performance needs to be combined with the uncertainties in the reference dataset (see Sect.~\ref{errordescription}). In Fig.~\ref{fig:disterror}, we show the MLP error distribution obtained on the test subsample (with no other error contribution, i.e. the difference with respect to SoS-Spectro).

In order to characterize the model performance, we report various error determinations in Table \ref{tab:statsres}. We observe that the results on both the SDSS and SM show only modest biases of $\simeq$50\,K in T$_{\rm{eff}}$, $\simeq$0.1\,dex in log\,$g$, and less than 0.1\,dex in [Fe/H]. 
The fact that the mean and the $\sigma$ are slightly higher than the median and the MAD is due to a long tail of rare, but high errors, as can be seen in Fig. \ref{fig:disterror}.

\begin{figure*}[t]
   \centering
   \includegraphics[width=1.\textwidth]{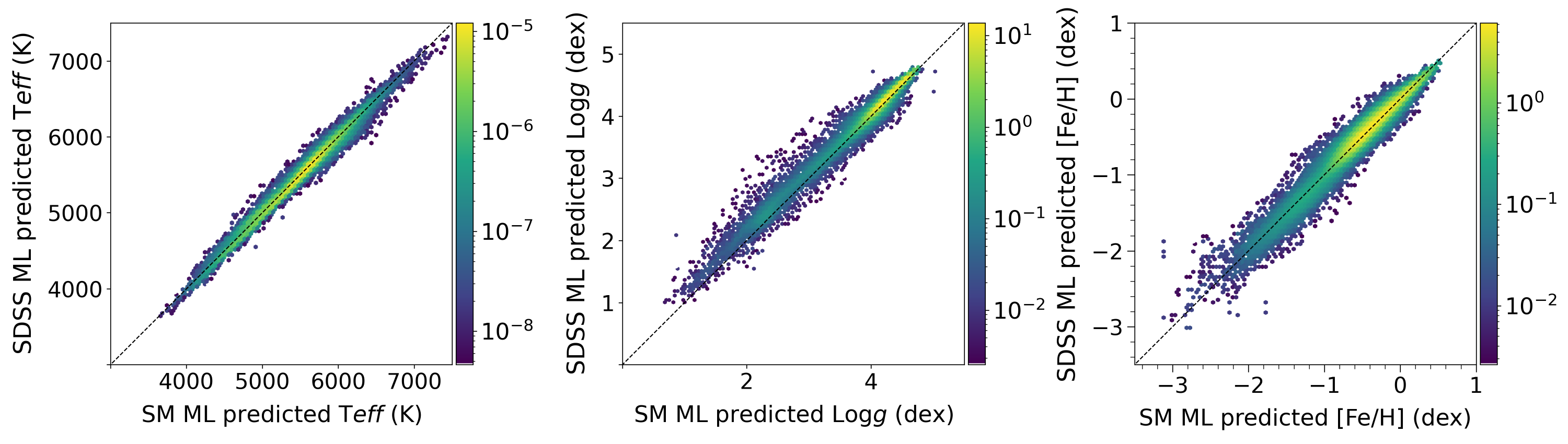}
      \caption{Comparison between the ML predictions obtained on the intersection between SDSS and SM training+test sample and SM sample (55696 stars). Lighter color corresponds to higher data density per pixel.}
   \label{fig:resultintersect}
   \end{figure*}

In figure \ref{fig:resultintersect}, we show the comparison between the ML predictions on the stars in common between the SM and SDSS training and test datasets (55696 stars). The models used for each set come from a different training on slightly different parameters, so the excellent agreement between the two predictions is a proof that the method is robust enough to be preliminarily trusted. We find median differences of 26\,K in T$_{\rm{eff}}$, 0.03\,dex in log\,$g$, and 0.04\,dex in [Fe/H].

\section{Machine Learning Survey of Surveys}
\label{sec:res}
After validating the performance of the method on the test sample, we run the trained model on both the SM and SDSS full datasets (i.e., not just the stars matched against SoS-Spectro), each accounting for around 9-10 million stars after applying the relevant filtering (see Sects.~\ref{sec:sdss} and \ref{sec:sm}).

\subsection{Error estimation}
\label{errordescription}

In Table \ref{tab:statsres}, we reported the median error obtained on the test subsample by comparing the ML-predicted values with the SoS-Spectro reference ones. In the following we will refer to this error component as $e_{\rm{train}}$, which represents the accuracy of the ML method.

In order to test the repeatability of the model predictions against the choice of the training sample, for each of the SDSS and SM datasets, we realized 10 different trainings of the model starting from different random choices of the training, validation and test samples. For each parameter, we then computed the MAD of the predictions for each star in the full set, obtained with the 10 differently trained models, along with the median prediction. We use our median prediction as our actual prediction, thus mitigating the impact of potential random outliers, and we use the MAD as a repeatability (precision) error component, hereafter referred to as $e_{\rm{rep}}$.

Our final error estimate, for each parameter and each star in the SM and SDSS datasets, is the result of the combination of the two previously identified error components ($e_{\rm{train}}$ and $e_{\rm{rep}}$), which are considered independent and thus summed in quadrature:
\begin{equation}
    \label{eq:error}
    e_{\mathrm{ML}} = \sqrt{e_{\mathrm{train}}^2 + e_{\mathrm{rep}}^2}
\end{equation}
This is the error that we include in the final catalog (Sect.~\ref{sec:cat}). In Table \ref{tab:finalerrtable}, we report its mean and median value for each parameter and each dataset considered in this work.
The Kiel diagrams of the two sets of predictions for the SDSS and SM datasets, colored by the estimated errors, are shown and discussed in Appendix~\ref{appendix:separate_survey_results}, together for the respective error distributions. 

\begin{table}
    \centering
    \caption{Final error estimate $e_{\rm{ML}}$ (Eq. \ref{eq:error}) on the full datasets.}
    \label{tab:finalerrtable}
    \begin{tabular}{lccc}
    \hline
    \hline
        & T$_{\rm{eff}}$ & log\,$g$ &  [Fe/H] \\
        & (K) & (dex) & (dex) \\
        \hline
        \multicolumn{4}{c}{{\em SDSS full sample:}} \\
        $e_{\rm{ML}}$ (mean) & 59 & 0.09 & 0.09\\
        $e_{\rm{ML}}$ (median) & 52 & 0.08 & 0.08\\
        \hline
        \multicolumn{4}{c}{{\em SM full sample:}} \\
        $e_{\rm{ML}}$ (mean) & 63 & 0.08 & 0.08\\
        $e_{\rm{ML}}$ (median) & 56 & 0.07 & 0.07\\
        \hline
        \multicolumn{4}{c}{{\em Combined final sample:}} \\
        $e_{\rm{ML}}$ (mean) & 61 & 0.08 & 0.08 \\
        $e_{\rm{ML}}$ (median) & 54 & 0.07 & 0.08 \\
        $\sigma$ & 21 & 0.02 & 0.02 \\
        MAD & 5.6 & 0.01 & 0.01 \\
        \hline
    \end{tabular}
\end{table}

\begin{table}
\caption{The second release of the Survey of Surveys, containing both the SoS-Spectro and the SoS-ML catalogs.}
\label{tab:cat}
\centering                         
\begin{tabular}{lcl}        
\hline\hline                
Column              & Units & Description \\
\hline  
SoSid & & Unique SoS identifier (DR2)\\
{\em Gaia} DR2 id   &          & {\tt source\_id} from {\em Gaia} DR2 \\
{\em Gaia} DR3 id   &          &  {\tt source\_id} from {\em Gaia} DR3 \\
SDSS objectId   &          &  Source id from SDSS \\
SM objectId   &          &  Source id from SM \\
RA                  & (deg) & RA from {\em Gaia} DR3 \\
Dec                 & (deg) & Dec from {\em Gaia} DR3 \\
spec\_T${\rm{eff}}$ & (K) & SoS-Spectro$^a$ temperature \\
spec\_err\_T${\rm{eff}}$ & (K) & Error on spec\_T$_{\rm{eff}}$$^a$ \\
spec\_log\_g & (dex) & SoS-Spectro$^a$ gravity \\
spec\_err\_log\_g & (dex) & Error on spec\_log\_$g$ \\
spec\_$[$Fe/H$]$ & (dex) & SoS-Spectro$^a$ metallicity \\
spec\_err\_$[$Fe/H$]$ & (dex) & Error on spec\_$[$Fe/H$]$ \\
ml\_T${\rm{eff}}$ & (K) & Temperature from ML$^b$ \\
ml\_err\_T${\rm{eff}}$ & (K) & Error on  ml\_T$_{\rm{eff}}$ \\
ml\_log\_g & (dex) & Gravity from ML$^b$ \\
ml\_err\_log\_g & (dex) & Error on ml\_log\_$g$ \\
ml\_$[$Fe/H$]$ & (dex) & Metallicity from ML$^b$ \\
ml\_err\_$[$Fe/H$]$ & (dex) & Error on ml\_$[$Fe/H$]$ \\
source & & source catalogue ('SDSS', \\
       & & 'SM', or 'SDSS+SM') \\
SDSS\_train\_set & (bool) & Star in SDSS training set \\
SM\_train\_set & (bool) & Star in SM training set \\
SDSS\_train\_area & (bool) & Star in SDSS training area \\
SM\_train\_area & (bool) & Star in SM training area \\

\hline 
\multicolumn{3}{l}{$^a$ see Sect.~\ref{sec:specsos} for details.}\\
\multicolumn{3}{l}{$^b$ see Sect.~\ref{sec:res} for details.}\\
\end{tabular}
\end{table}

   \begin{figure*}[t!]
   \centering
   \includegraphics[width=\textwidth]{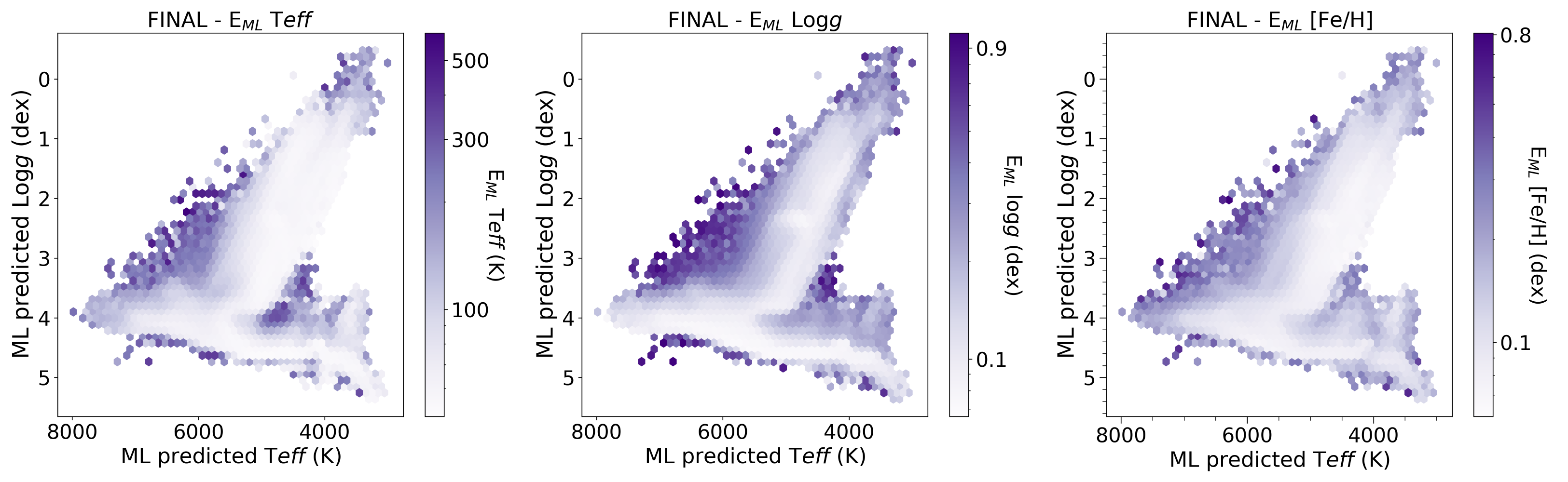}
       \caption{Kiel Diagram for the final catalog, colored with the estimated errors on the three parameters. From left to right: T$_{\rm{eff}}$, log\,$g$ and [Fe/H]. The diagram is divided in small hexagonal bins and the color represents the average of the error inside the bin.}
   \label{fig:finalkiel}
   \end{figure*}

\subsection{Final catalog}
\label{sec:cat}

To compose the final catalog of ML-predicted parameters, we merged the SM and SDSS datasets by averaging the predictions obtained on stars in common, which are compatible with each other (see Fig. \ref{fig:resultintersect}). We applied the following filters on the final ML-predicted parameters and errors:
\begin{itemize}
    \item T$_{\rm{eff}}$ < 3000 K or T$_{\rm{eff}}$ > 8000 K
    \item log\,$g$ < --0.5 dex or log\,$g$ > 5.5 dex
    \item $[$Fe/H$]$ $<$ --3.5 dex or [Fe/H] > 1.0 dex
    \item $e_{\rm{ML}}$ on T$_{\rm{eff}}$ > 10\%
    \item $e_{\rm{ML}}$ on log\,$g$ > 1.0 dex
    \item $e_{\rm{ML}}$ on [Fe/H] > 1.0 dex
\end{itemize}
The final catalog contains around 19 million stars obtained with ML, and only 1557 were removed using the above criteria. To this catalog we added, in separate columns (see Tab. \ref{tab:cat}), the SoS-Spectro measurements, including the stars from SoS-spectro which did not crossmatch to Gaia + SDSS or to Gaia + SM, or where we could not apply the ML algorithm due to the quality constraints on the input parameters described in the selection criteria in Sec. \ref{refdatasets}, which account for around 3.75 million stars, and do not have a corresponding ML estimate. The final catalog composed in this way contains around 23 million stars.
The catalog contains a newly defined SoSid, built combining a healpix index with a running number, with the aim of ensuring the id uniqueness throughout the SoS project.
In Table \ref{tab:finalerrtable}, we show the statistical properties of the final estimated ML errors for the three parameters on the merged catalog. We note that the spread on the errors is very small, as indicated by the $\sigma$ and MAD, and the mean and median values are in general compatible with those typically found in spectroscopic surveys.
We apply the ML model to stars which may be not enclosed in the training set (see Fig.~\ref{KielA} in ~\ref{appendix:separate_survey_results}), thus we decided to evaluate if the input parameters used to compute the ML values of the predicted quantities are compatible with the region spanned by the training set. We computed a Probability Density Function (PDF) with the treeKDE algorithm \citep{treekde}, which is extremely fast on large datasets, over the full dimensionality of the training parameter spaces (SDSS and SM), and we added a boolean flag (``SDSS\_train\_area'' or ``SM\_train\_area'') set to ``true'' if the input features fall into the 99\% threshold of the region defined by the respective PDF.

In Fig. \ref{fig:finalkiel}, we show the Kiel diagram for the final combined catalog, colored with the errors on the three parameters. The errors appear very similarly distributed to the ones of the SDSS and SM catalogs seen separately (Fig.\ref{KielA}) but the error range is noticeably smaller after the quality cuts described above. The areas with larger errors are generally at the margins of the main body of the distributions, which are often outside of the parameter range covered by the reference catalog, the SoS-Spectro. 

Notably, higher errors are also visible: {\em (i)} in the high temperature, low gravity region containing the Hertzsprung gap, the extended horizontal branch, and yellow stragglers; {\em (ii)} in the cool dwarfs and pre-main sequence region to the lower right, which is known to be difficult to analyze accurately; and {\em (iii)} immediately below the main sequence, at any temperature, marking a region where likely binaries composed of a main sequence and a white dwarf star could lie, as well as sub-dwarfs in general.

The catalog presented in Table~\ref{tab:cat} combines the SoS-Spectro and SoS-ML catalogs and constitutes the second release of the SoS.

\section{Science validation}
\label{sec:val}

We carried out several tests to validate our results, to understand the strengths and weaknesses of our catalog, and to compare with other literature sources. We discuss here the comparisons with other ML methods and the validation with star clusters. Other validation tests are presented in Appendix~\ref{sec:add_checks}.

   \begin{figure*}[t!]
   \centering
   \includegraphics[width=\textwidth]{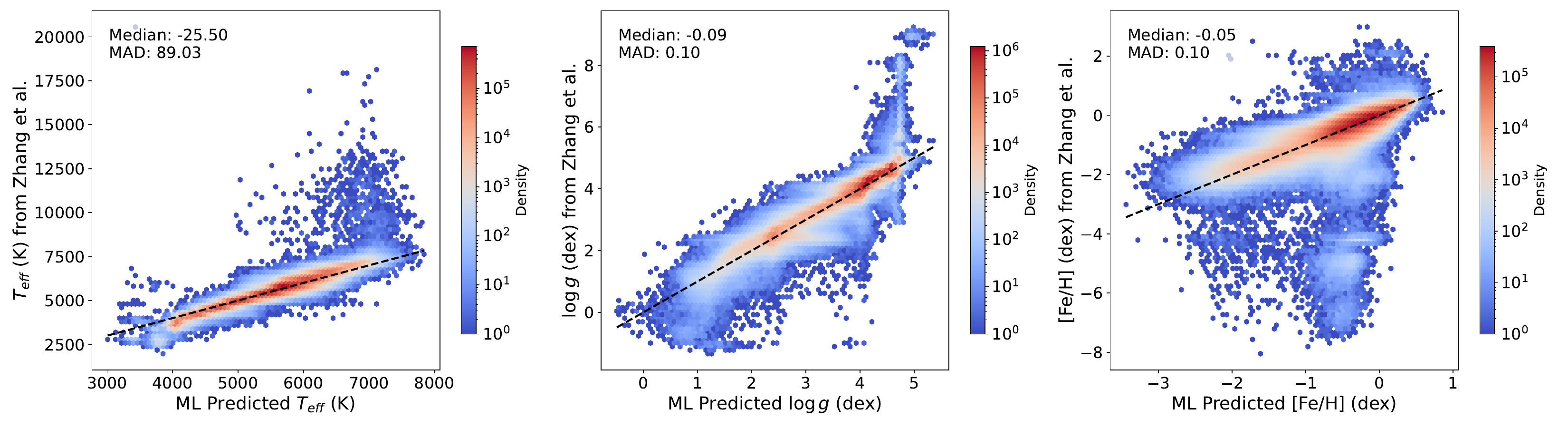}
   \includegraphics[width=\textwidth]{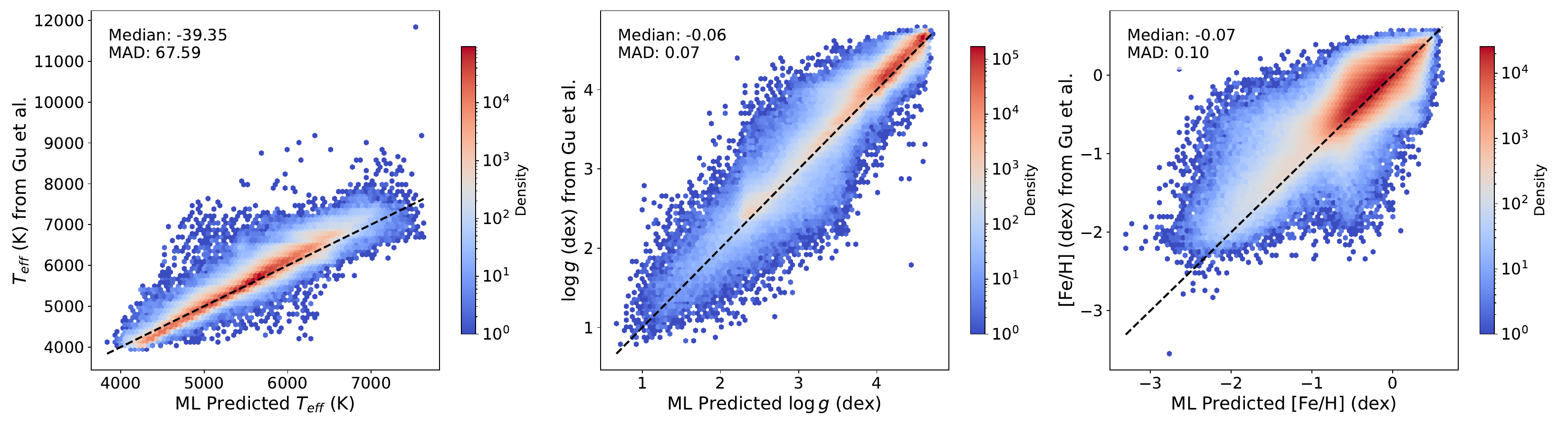}
   \includegraphics[width=\textwidth]{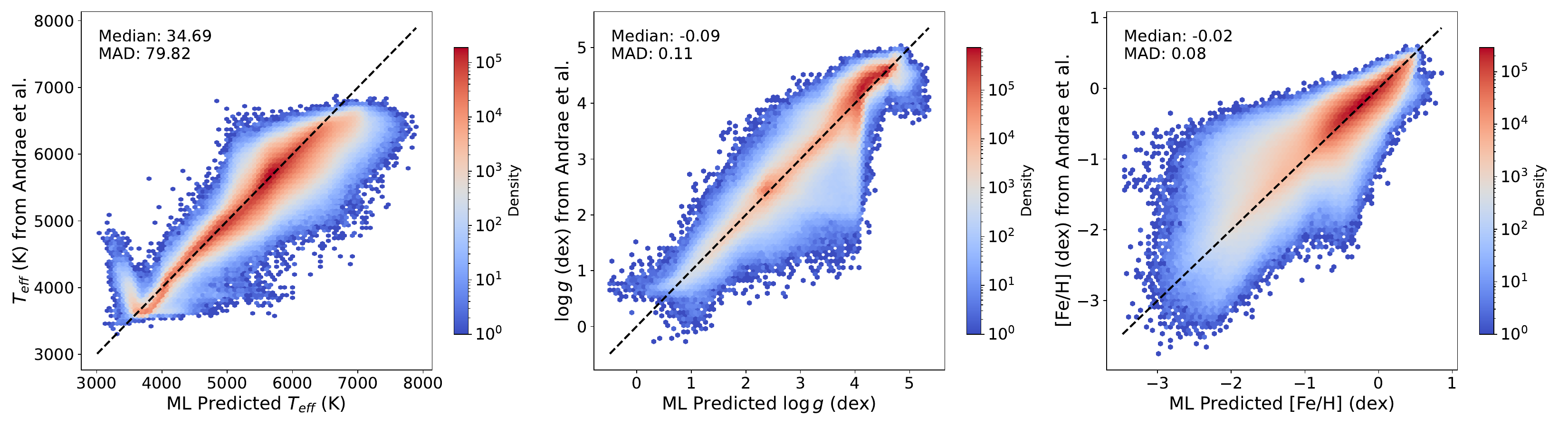}
      \caption{Comparison of atmospheric parameters predicted by our ML approach (abscissae) and literature ML catalogs (ordinates). The left plots show T$_{\rm{eff}}$ comparisons, the middle ones log\,$g$, and the right ones [Fe/H]. The top row shows the comparison with \citet{Zhang2023}, the middle row with \citet{gu25}, and the bottom row with \citet{andrae23}. The dashed line represents the 1:1 relationship.}
   \label{fig:litml}
   \end{figure*}

\subsection{Comparison with Zhang et al.} 

\citet{Zhang2023} employed machine learning to obtain atmospheric parameters from Gaia XP spectra. They utilized a data-driven forward model trained on LAMOST and incorporated near-infrared photometry to reduce degeneracies in parameter estimation. The approach is based on modeling the Gaia XP spectra with known astrophysical parameters and then inferring the parameters for all Gaia XP spectra. 

In the top row of Fig.~\ref{fig:litml}, we compare the parameters obtained in this work with those by \citet{Zhang2023}, focusing on the stars for which they report a confidence level higher than 0.5.
For the vast majority of stars, we find excellent agreement, with very small trends and biases. However, there are long tails of discrepant determinations in all three parameters, which concern a small fraction of the stars in common. 
For log\,$g$, some objects in \citet{Zhang2023} display values that extend up to 8, which is significantly higher than the typical range for main-sequence stars and giants. Similarly, we observe discrepancies in [Fe/H] values, which reach values as low as --7\,dex in \citet{Zhang2023}.  Metallicities lower than -4 are not typical for the majority of stars and would be considered extreme, corresponding to rare, very old stars or, most likely, artifacts in the data, as the LAMOST survey does not have metallicity values lower than -2.5. Stars exhibiting extreme log\,$g$ and [Fe/H] values are the same objects, supporting the idea that these may be artifacts in the \cite{Zhang2023} results. However, this cannot be said with confidence about the major offset in effective temperatures. Most stars in the hot star subset identified in their work show good agreement with our results for other parameters. It is worth noting here that the subset of hot stars identified by \cite{Zhang2023} is concentrated near the Galactic plane. Since effective temperature estimates depend on accurate extinction corrections, it is difficult to determine which set of T$_{\rm{eff}}$ values is more reliable in this case.  

\subsection{Comparison with Gu et al.} 

The second release of the SAGES database \citep{gu25}
presents an ML derivation of stellar parameters for over 21 million stars, based on a vast compilation of data obtained by integrating spectroscopic estimates and high-resolution data from the literature with photometric observations from SAGES DR1 \citep{fan23}, Gaia DR3 \citep{gdr3}, All-WISE \citep{cutri21}, 2MASS \citep{cutri03}, and GALEX \citep{bianchi14}. They employ an ML approach based on the random forest algorithm to infer T$_{\rm{eff}}$, log\,$g$, and metallicity. As labels, the data from LAMOST DR10 \citep{lamost} and APOGEE DR17 \citep{apogee17} was used, augmented with the data from PASTEL \citep{soubiran16} and RAVE DR5 \citep{rave5}.

Given the similarity in methodology and size of the resulting catalog with ours, the comparison with the SAGE catalog is particularly relevant; it is shown in the middle row of Fig.\ref{fig:litml} and shows excellent agreement. There are no significant groups of outliers, unlike in the comparison with \citet{Zhang2023}. We only noticed a small trend in T$_{\rm{eff}}$, which is qualitatively similar to the one in the comparison with \citet{Zhang2023}, but completely different from the one in the comparison with \citet{andrae23}. The observed trend appears well within the combined uncertainties reported in the catalogs ($<$150\,K at the extremes), but it would be interesting to further explore this feature in the future.

\subsection{Comparison with Andrae et al.} 

In our ML approach, we used the metallicity [Fe/H] from \citet{andrae23} as an input parameter. Our choice was motivated by the far better comparison of their metallicities with the SoS-Spectro and PASTEL, with respect to the {\em Gaia} DR3 GSP-Phot metallicities. We did not use their $T_{\rm eff}$ and $\log g$ values, therefore a comparison is still relevant. \cite{andrae23} applied the XGBoost machine-learning algorithm to publicly available {\em Gaia} XP spectra, training their model on stellar parameters from the APOGEE survey, complemented with a set of very metal-poor stars.  Their catalog is vast compared to ours, comprising 175 million stars.

\begin{figure*}[t]
\centering
\begin{minipage}{\textwidth}
   \centering
   \includegraphics[width=\textwidth]{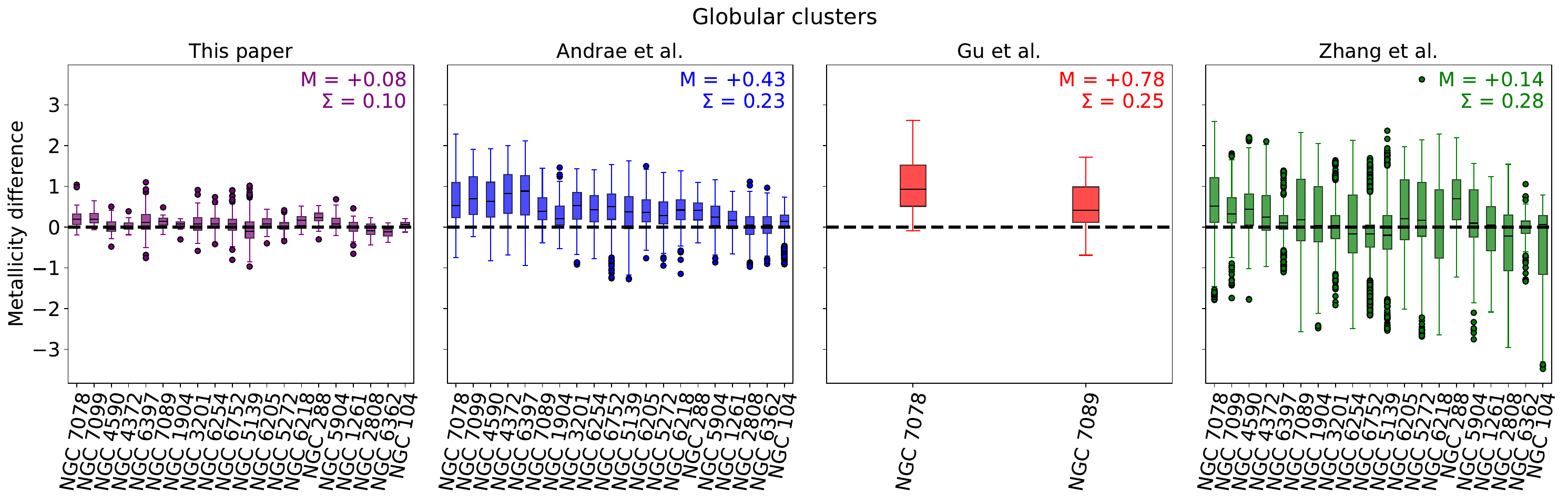}
\end{minipage}
\\
\vspace{0.5cm}
\begin{minipage}{\textwidth}
   \centering
   \includegraphics[width=\textwidth]{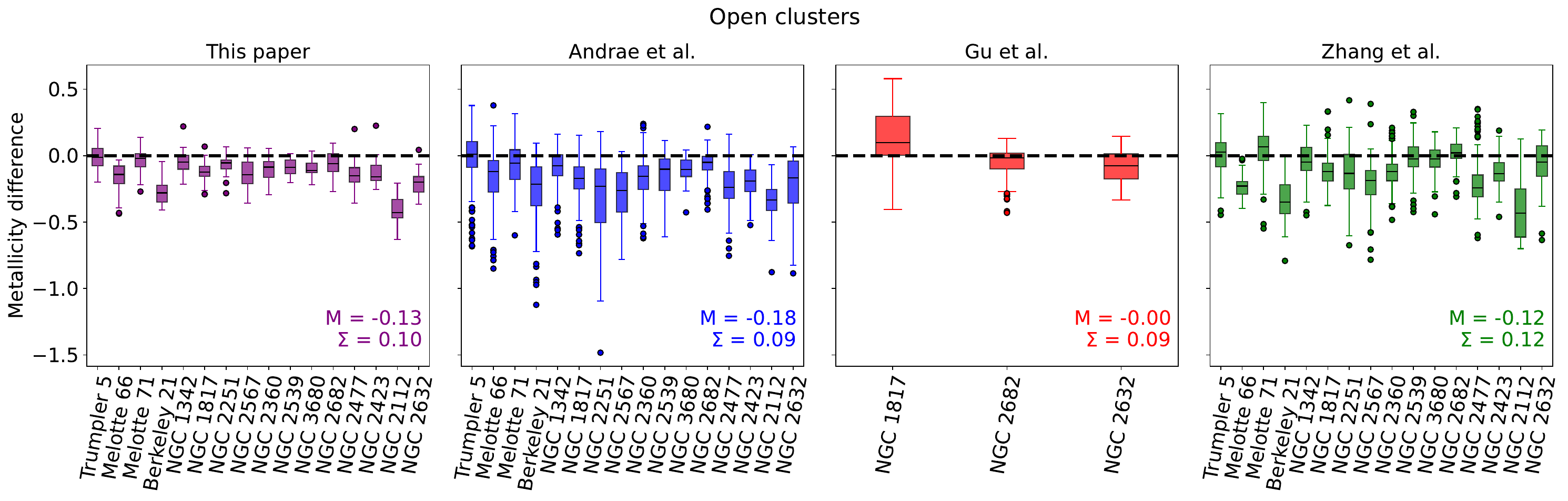}
\end{minipage}
\caption{Comparison of metallicity predictions for globular (top row) and open clusters (bottom row). Our results are shown in purple in the leftmost panels, the ones by \citet{andrae23} in blue in the center-left panels, the ones by \citet{gu25} in red in the center-right panels, and the ones by \citet{Zhang2023} in green in the rightmost panels. The boxplots show the metallicity differences with respect to the \citet{Harris1996} and \citet{Netopil2016} values for selected clusters. The boxes represent the interquartile range (IQR), the central lines the median, and the whiskers extend to 1.5 times the IQR. Outliers are marked as individual points. The mean metallicity difference (M) and standard deviation ($\Sigma$) are annotated in each panel. The dashed line at zero indicates perfect agreement with the literature values (zero-line). Clusters are sorted by ascending metallicity.} 
\label{fig:clusters_metallicity_summary}
\end{figure*}

The bottom row of Fig.~\ref{fig:litml} shows the comparison. While the [Fe/H] values agree extremely well, as expected since we used them as input parameters to the ML prediction, we observe a large spread. Combined with our favorable comparison with star clusters (Sect.~\ref{clustervalid}), this shows that indeed, by building upon the work by \citet{andrae23}, we were able to obtain improved predictions with our approach. On the other hand, we observe striking patterns in the comparison with $T_{\rm eff}$ and $\log g$, which are not present in the comparisons with \citet{Zhang2023} and \citet{gu25}. Interestingly, \citet{gu25} observe the same patterns in the comparison of their $\log g$ values with those from \citet{andrae23}, which points to features in the \citet{andrae23} catalog rather than in ours.

\subsection{Validation on star clusters}
\label{clustervalid}

To further test our metallicity predictions, we validated our results using a sample of globular and open clusters. For globular clusters, the metallicities used for validation were taken from \citet{Harris1996}, while we selected probable members using \citet{VasilievBaumgardt2021} with a membership probability threshold of 0.8. We focused on 20 globular clusters with more than 20 members having ML predictions in our catalog. For open clusters, we used high-quality metallicities from \citet{Netopil2016} and we selected members using  \citet{HuntReffert2024}. As with the globular clusters, we focused on 16 open clusters having more than 20 members in our ML sample.
In Fig.~\ref{fig:clusters_metallicity_summary}, we extended the comparison to the other ML catalogues examined in the previous sections, using the same clusters, reference metallicities, and members selection procedure.

For globular clusters, which are notoriously difficult to parametrize correctly with ML methods or even in the typical spectroscopic surveys, our predictions are excellent and they appear to have smaller offsets from \citet{Harris1996} and smaller spreads compared to those of the other ML methods. Additionally, our predictions are equally good at all metallicity, unlike in the case of other ML methods. This proves that our approach, of building upon previous work \citep[in this case by][]{andrae23} to improve on it, rather than starting from scratch, does really help in improving predictions at the low metallicity end. We also believe that our recalibration of the SoS-Spectro on PASTEL played an important role, precisely by improving the [Fe/H] reference values for metal-poor giants, and thus globular clusters. 

For open clusters, we observe an underestimation in the predictions of all ML methods, except for \citet{gu25}. The offset is of the order of --0.1\,dex or a bit more, depending on the method.  Despite having better mean values, \citet{gu25} shows a larger spread in their predictions within each cluster. A detailed cluster-by-cluster, star-by-star comparison is presented in Appendix~\ref{sec:per_cluster}. Our approach appears to perform similarly to the other methods at Solar metallicity, with a small improvement on the spread within each cluster compared to other ML methods, and also a small improvement on the zeropoint compared to our input dataset, \citet{andrae23}. Concerning the observed offset with \citet{Netopil2016}, the extremely similar behavior among the different ML methods suggests the possibility that the observed systematics are in the \citet{Netopil2016} catalog. We thus tried to compare to other literature sources such as \citet{dias21} or \citet{kharchenko13}, which is a compilation of metallicities from other sources, but we obtained very similar --0.1\,dex offsets, so the origin of the offsets remains for the moment unexplained.

\section{Conclusions}
\label{sec:con}

In this paper, we presented the second release of the SoS catalog containing atmospheric parameters, namely T$_{\rm{eff}}$, log\,$g$, and [Fe/H], for about 23 million stars, with estimated formal errors of about 50\,K for T$_{\rm{eff}}$, and around 0.07-0.08\,dex for log\,$g$ and [Fe/H]. The catalog is composed of two parts: SoS-Spectro and SoS-ML. The SoS-Spectro catalog (Sect.~\ref{sec:specsos}) is the result of a simple homogenization procedure applied to the same five spectroscopic surveys of \citet{tsantaki22}, recalibrated on the high-resolution spectroscopy analysis collected by \citet{soubiran16} in PASTEL.

The SoS-ML catalog (Sect.~\ref{sec:res}) is the result of a ML method application to enhance previous estimates of stellar parameters, i.e., to improve their precision and accuracy. In particular, we relied on T$_{\rm{eff}}$ and log\,$g$, estimates by {\em Gaia} DR3 and [Fe/H] estimates by \citet{andrae23}. We used large photometry surveys, such as SDSS and SM to refine the parameters. This was made possible also by leveraging the large {\em Gaia} DR3 dataset with precise astrometry and photometry, which was also key to our accurate cross-match between catalogs. In our work, we selected a very simple Multi-layer perceptron architecture (Sect.~\ref{sec:nn}) that yielded extremely good results, even when comparing with recent literature (Sect.~\ref{sec:val}). Compared to previous efforts in deriving stellar parameters with ML, our method has two main differences. The first key ingredient is the SoS-Spectro reference catalog, which is recalibrated on high-resolution spectroscopy: our model was trained on the SoS values to accurately reproduce them from the photometric inputs. The second is that we do not try to predict parameters starting from scratch, but we build on previous estimates by {\em Gaia} DR3 and \citet{andrae23}.

To summarize our validation results (Sect.~\ref{sec:val} and App.~\ref{sec:add_checks}), we compare extremely well with other ML catalogs in the literature, for the vast majority of stars in common. However, we do not show some of the samples of spurious determinations that are visible in other catalogs. Additionally, our results show a smaller spread, i.e., our internal errors (precision) are smaller. Our catalog size is comparable with some ML catalogs in the literature \citep[e.g.,][]{Zhang2023,gu25}, but is one order of magnitude smaller than \citet{andrae23} and two orders of magnitude smaller than {\em Gaia} DR3. When comparing with star clusters, we perform similarly to, or slightly better than, other ML methods at Solar metallicity, but there is a striking improvement in the globular cluster metallicity range, both in accuracy and precision. We also tried to improve on the problems that ML methods face at even lower metallicities, by augmenting our training set with a few hundred very metal-poor stars ([Fe/H]$\,\lesssim$\,--2\,dex, Appendix~\ref{app:vmp}), but this only produced a slight improvement at all metallicities, without actually solving the problem. 

In future work, we will try to address the problems at very low metallicities but also work on increasing the sample sizes by including more photometric and spectroscopic surveys. Adding abundance ratios to the list of parameters would surely be a worthwhile goal. It is also vital -- in general -- to refine the extinction and distance inputs to improve on the quality and quantity of the predictions. Finally, several additional improvements can be achieved by using more sophisticated algorithmic approaches that make use of pre-categorization, clustering, and feature enhancement. 

\section{Data availability}
The SoS DR2 catalog is available electronically at CDS\footnote{\url{https://vizier.cds.unistra.fr/viz-bin/VizieR}} and on the SoS portal at the Space Science Data Center\footnote{\url{https://gaiaportal.ssdc.asi.it/SoS2}}.

\begin{acknowledgements}

People. We acknowledge interesting exchanges of ideas with our colleagues: G.Battaglia, S.~Fabbro, I.~Gonzalez Rivera, G.~Kordopatis, M.~Valentini.\\
Funding. Funded by the European Union (ERC-2022-AdG, {\em "StarDance: the non-canonical evolution of stars in clusters"}, Grant Agreement 101093572, PI: E. Pancino). Views and opinions expressed are however those of the author(s) only and do not necessarily reflect those of the European Union or the European Research Council. Neither the European Union nor the granting authority can be held responsible for them. We also acknowledge the financial support to this research by INAF, through the Mainstream Grant {\em “Chemo-dynamics of globular clusters: the Gaia revolution”} (n. 1.05.01.86.22, P.I. E.~Pancino). We are thankful for the
team meetings at the International Space Science Institute
(Bern) for fruitful discussions and were supported by the ISSI
International Team project "AsteroSHOP: large Spectroscopic surveys HOmogenisation Program" (PI: G.~Thomas). EP acknowledges financial support from PRIN-MIUR-22: CHRONOS: adjusting the clock(s) to unveil the CHRONO-chemo-dynamical Structure of the Galaxy” (PI: S.~Cassisi) funded by the European Union -- Next Generation EU. MT thanks INAF for the Large Grant EPOCH and the Mini-Grant PILOT (1.05.23.04.02). PMM and SM acknowledge financial support from the ASI-INAF agreement n. 2022-14-HH.0. GFT acknowledges support from the Agencia Estatal de Investigaci\'on 
del Ministerio de Ciencia en Innovaci\'on (AEI-MICIN) and the European Regional Development Fund (ERDF) under grant numbers PID2020-118778GB-I00/10.13039/501100011033 and PID2023-150319NB-C21/C22 and the AEI under grant number CEX2019-000920-S. GG acknowledges support by Deutsche Forschungs-gemeinschaft (DFG, German Research Foundation) – project-IDs: eBer-22-59652 (GU 2240/1-1 "Galactic Archaeology with Convolutional Neural-Networks: Realising the potential of Gaia and 4MOST"). This project has received funding from the European Research Council (ERC) under the European Union’s Horizon 2020 research and innovation programme (Grant agreement No. 949173). FG gratefully acknowledges support from the French National Research Agency (ANR) funded projects ``MWDisc'' (ANR-20-CE31-0004) and ``Pristine'' (ANR-18-CE31-0017)\\
Data. This work has made use of data from the European Space Agency (ESA) mission Gaia (\url{https://www.cosmos.esa.int/gaia}), processed by the Gaia Data Processing and Analysis Consortium (DPAC, \url{https://www.cosmos.esa.int/web/gaia/dpac/consortium}). Funding for the DPAC has been provided by national institutions, in particular the institutions participating in the Gaia Multilateral Agreement. 
Funding for the Sloan Digital Sky Survey IV has been provided by the Alfred P. Sloan Foundation, the U.S. Department of Energy Office of Science, and the Participating Institutions. SDSS-IV acknowledges support and resources from the Center for High Performance Computing at the University of Utah.
The national facility capability for SkyMapper has been funded through ARC LIEF grant LE130100104 from the Australian Research Council, awarded to the University of Sydney, the Australian National University, Swinburne University of Technology, the University of Queensland, the University of Western Australia, the University of Melbourne, Curtin University of Technology, Monash University and the Australian Astronomical Observatory.\\
Software. Most of the plotting and data analysis was carried out using R \citep{R,data.table} and Python \citep{keras,tensorflow}.This research has made use of the SIMBAD database \citep{simbad} and the VizieR catalogue access tool \citep{vizier}, both operated at CDS, Strasbourg, France. Preliminary data exploration relied heavily on TopCat \citep{taylor05}. 

\end{acknowledgements}

\bibliographystyle{aa}
\bibliography{MLSoS}

\begin{appendix}

\section{Systematics among survey releases}
\label{sec:allmaps}

   \begin{figure*}[t!]
   \centering
   \includegraphics[width=0.95\textwidth]{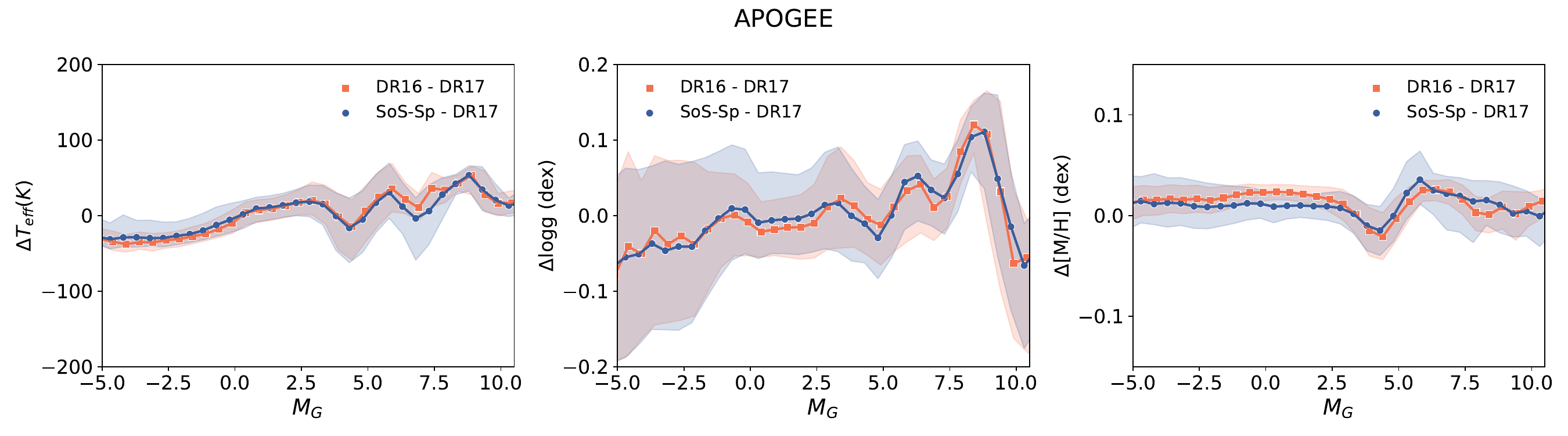}
   \includegraphics[width=0.95\textwidth]{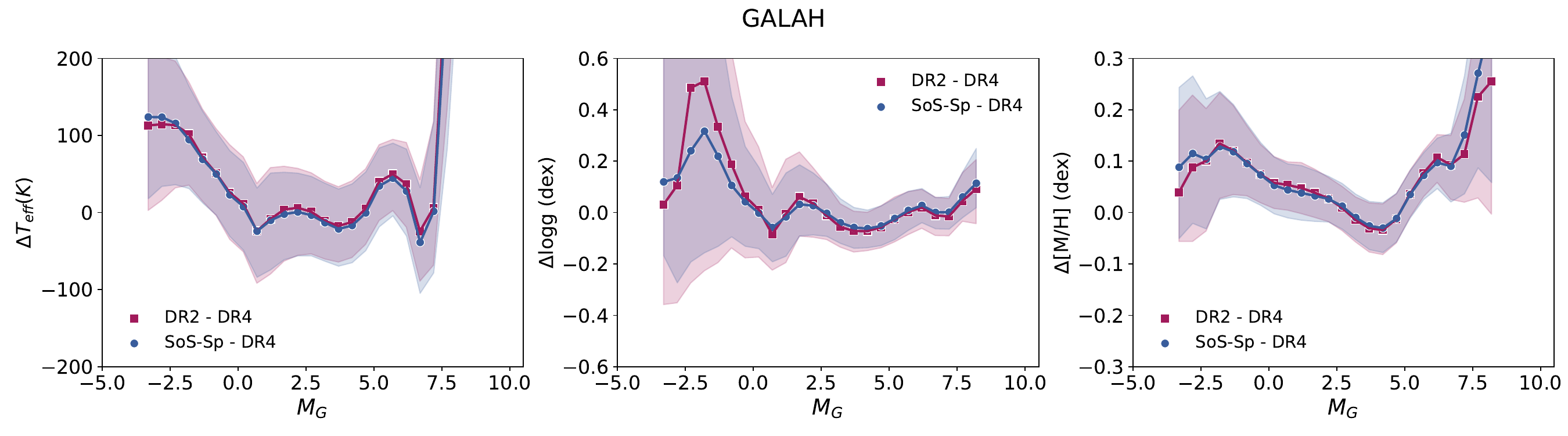}
   \includegraphics[width=0.95\textwidth]{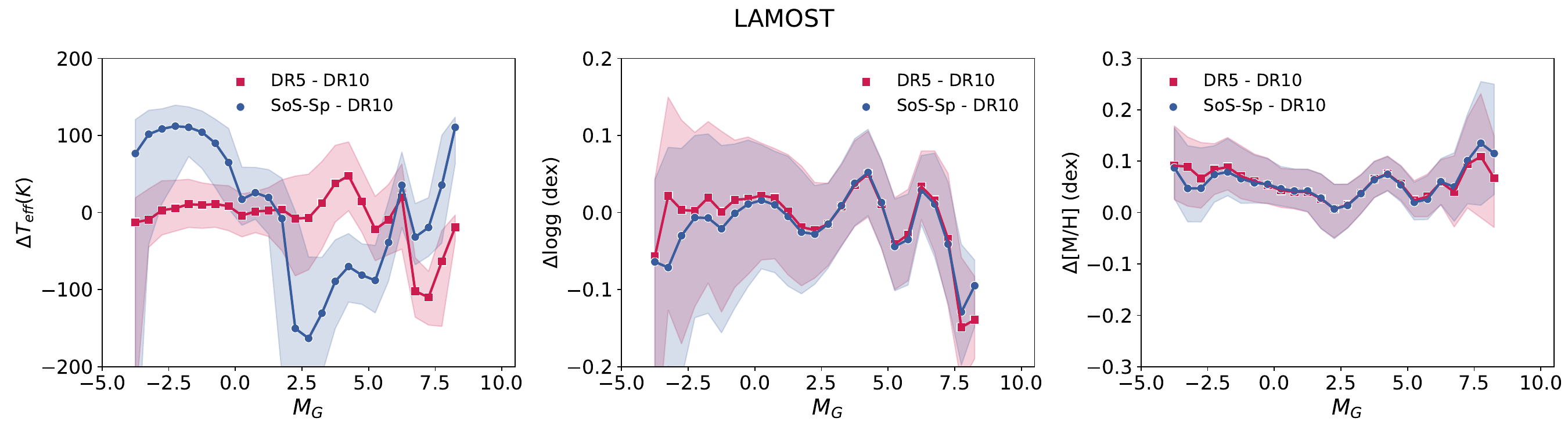}
      \caption{Differences in Teff , log g, and [Fe/H] between earlier and later data releases of three major spectroscopic surveys. \textit{Top}: APOGEE DR16 vs DR17 (orange) and SoS-Spectro vs DR17 (blue); \textit{middle}: GALAH DR2 vs DR4 (magenta) and SoS-Spectro vs DR4 (blue); \textit{bottom}: LAMOST DR5 vs DR10 (pink) and SoS-Spectro vs DR10 (blue). Square markers denote internal survey updates, round markers show differences with SoS-Spectro, and shaded regions represent the interquartile range of residuals. These comparisons reveal that the SoS-Spectro discrepancies are of similar or smaller scale than the internal changes across survey versions.}
   \label{fig:sos_bias_galah_full}
   \end{figure*}

In this section, we present a detailed comparison of the SoS-Spectro parameters with the most recent updates from the APOGEE DR17, GALAH DR4, and LAMOST DR10 surveys. As previously mentioned, the SoS-Spectro parameters are based on earlier data releases, which can lead to systematic biases when compared with the most recent versions. To ensure a fair comparison with high-quality data, we apply the following selection criteria, as recommended by the survey authors:

\begin{itemize}
    \item APOGEE DR17: \texttt{aspcapflag} $\neq$ \texttt{star\_bad}
    \item GALAH DR4: \texttt{flag\_sp} = 0, \texttt{flag\_fe\_h} = 0, \texttt{snr\_px\_ccd3} $>$ 30
    \item LAMOST DR10: \texttt{snr} $>$ 30
\end{itemize}

Figure~\ref{fig:sos_bias_galah_full} illustrates the differences between the SoS-Spectro parameters and the latest survey releases. The plots also include comparisons between the original data releases used in SoS-Spectro and the latest versions of each survey. We found that the systematic differences appear as wavy-like patterns when plotted against the absolute $G$-band magnitude from {\em Gaia} DR3. The typical differences between SoS-Spectro parameters and those from APOGEE DR17, GALAH DR4, and LAMOST DR10 are summarized in Table~\ref{tab:survey_comparison}. These observed differences reflect substantial improvements in the latest APOGEE and GALAH datasets. APOGEE DR17 features a reworked ASPCAP pipeline, a new synthetic spectral grid including NLTE effects, and auxiliary analyses using alternative libraries. GALAH DR4 benefits from a more stable wavelength calibration (especially near CCD edges) and enhanced outlier detection, yielding more reliable spectroscopic parameters.

More in details, in the comparison with GALAH DR4, the data show increased variability, with wider interquartile range (IQR) compared to APOGEE, especially in $T_{\rm eff}$ and log$g$. Notably, greater scatter in log$g$ is observed among stars brighter than $M_{G} = 0$. Despite this variability, the median offsets remain small, indicating overall consistency between SoS-Spectro and GALAH parameters.

Finally, we compared SoS-Spectro parameters with LAMOST DR10. While log$g$ shows negligible median deviation, a systematic offset of 0.05 dex is found between [Fe/H] values in DR5 and DR10, indicating a shift in calibration. The offset in $T_{\rm eff}$ between SoS-Spectro and DR10 is also clear and cannot be solely explained by differences with DR5. This offset originates from the specific corrections applied to LAMOST DR5 temperatures during the construction of the SoS-Spectro catalog.

\begin{table}[t!]
\centering
\caption{Comparison of SoS-Spectro parameters with the latest versions of the surveys.}
\label{tab:survey_comparison}
\begin{tabular}{lcc}
\hline
\hline
Parameter & Median Difference & Interquartile Range (IQR) \\
\hline
\multicolumn{3}{c}{{\em APOGEE DR17:}} \\
$T_{\rm eff}$ (K) & 6.53 & 50.24 \\
log\,$g$ (dex) & 0.01 & 0.10 \\
{[}Fe/H{]} (dex) & 0.01 & 0.04 \\
\hline
\multicolumn{3}{c}{{\em GALAH DR4:}} \\
$T_{\rm eff}$ (K) & -2.18 & 111.84 \\
log\,$g$ (dex) & -0.03 & 0.20 \\
{[}Fe/H{]} (dex) & 0.01 & 0.11 \\
\hline
\multicolumn{3}{c}{{\em LAMOST DR10:}} \\
$T_{\rm eff}$ (K) & -65.03 & 154.86 \\
log\,$g$ (dex) & 0.01 & 0.13 \\
{[}Fe/H{]} (dex) & 0.05 & 0.09 \\
\hline
\end{tabular}
\tablefoot{Extreme outliers ($>$3\,IQR from the median) are excluded from APOGEE, affecting up to 1.5\% of the data.}
\end{table}

\section{Uncertainties in SDSS and SM}

\label{appendix:separate_survey_results}
The training and application datasets are composed of two distinct parts based on SDSS and SM photometry. While the main text presents the combined results for clarity, this appendix provides the separate results for each survey, offering a more detailed view of the model performance.

\begin{figure*}[t]
   \centering
   \includegraphics[width=1.\textwidth]{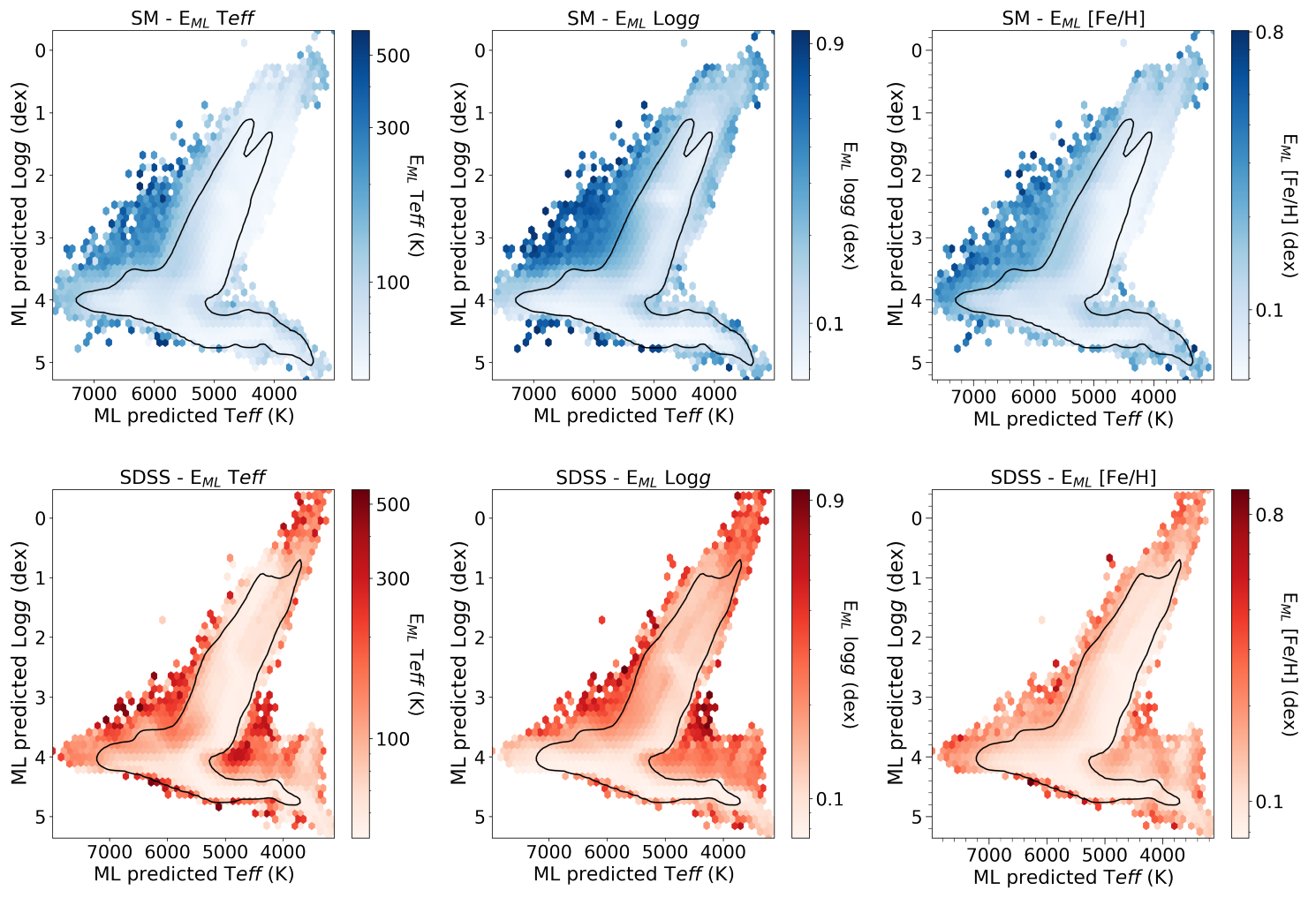}
      \caption{Kiel Diagram for the SM (above) and SDSS (below) full sample, colored with the estimated errors on the three parameters. From left to right: T$_{\rm{eff}}$, log\,$g$, and [Fe/H]. Hexagonal bins are colored based on the average of the errors inside the bin. The black line approximately encloses the region covered by the respective ``train\_area'' flags described in Sect.\ref{sec:cat}.}
   \label{KielA}
   \end{figure*}

In figure \ref{KielA}, we show the Kiel diagrams obtained on the full SDSS and SM catalogs, as selected in sections \ref{sec:sdss} and \ref{sec:sm} and colored with the estimated error $e_{\rm{ML}}$ on each parameter T$_{\rm{eff}}$, log\,$g$, and [Fe/H]. 
The error value is dominated by $e_{\rm{rep}}$ and thus indicate the repeatability of the prediction and the solidity of the model solution for the stars in that region of the parameter space. It is to be expected that the regions which have poor or no coverage within the training sample (indicated by the black line in Fig. \ref{KielA}) show the largest error.

\begin{figure*}[t]
   \centering
   \includegraphics[width=\textwidth]{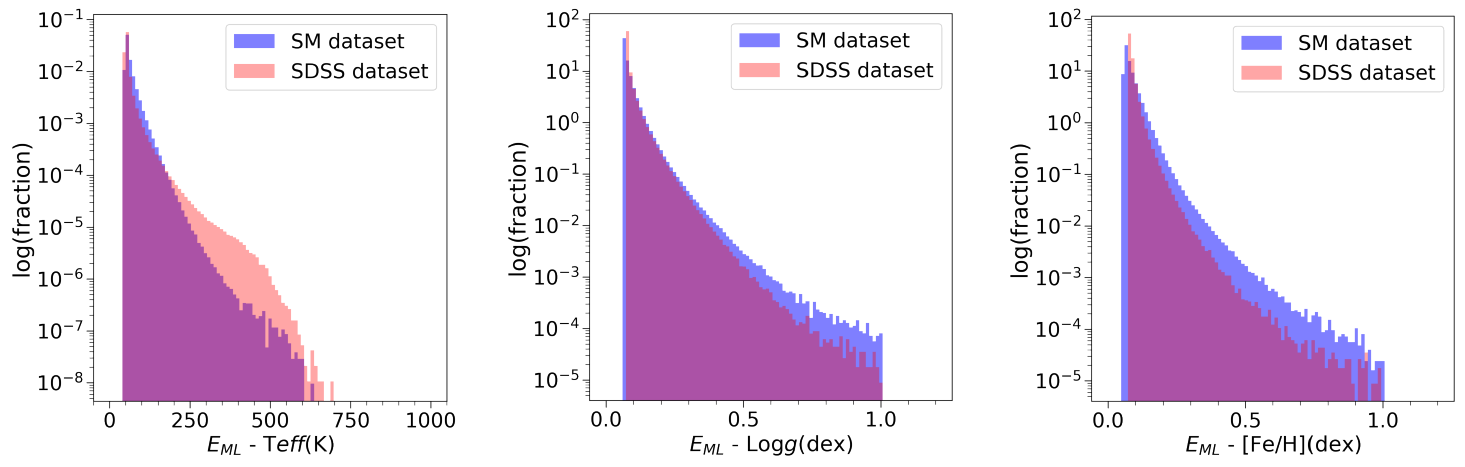}
       \caption{Distribution of $E_{ML}$ SM dataset (green) and the SDSS dataset (red). The y scale indicates the logarithm of the fraction of the dataset in each bin (total of 100 bins).}
   \label{fig:finalerrdist}
   \end{figure*}

In figure \ref{fig:finalerrdist}, we show the distributions of $e_{\rm{ML}}$ for the three parameters over each dataset. We notice that the distribution of [Fe/H] errors on the SM dataset has larger tails, which however are several order of magnitudes less populous than the distribution peak and thus do not affect significantly the global median or mean error.


\section{Additional validation}
\label{sec:add_checks}

\subsection{Very metal-poor stars}
\label{app:vmp}

\begin{figure*}[t]
   \centering
   \includegraphics[width=\textwidth]{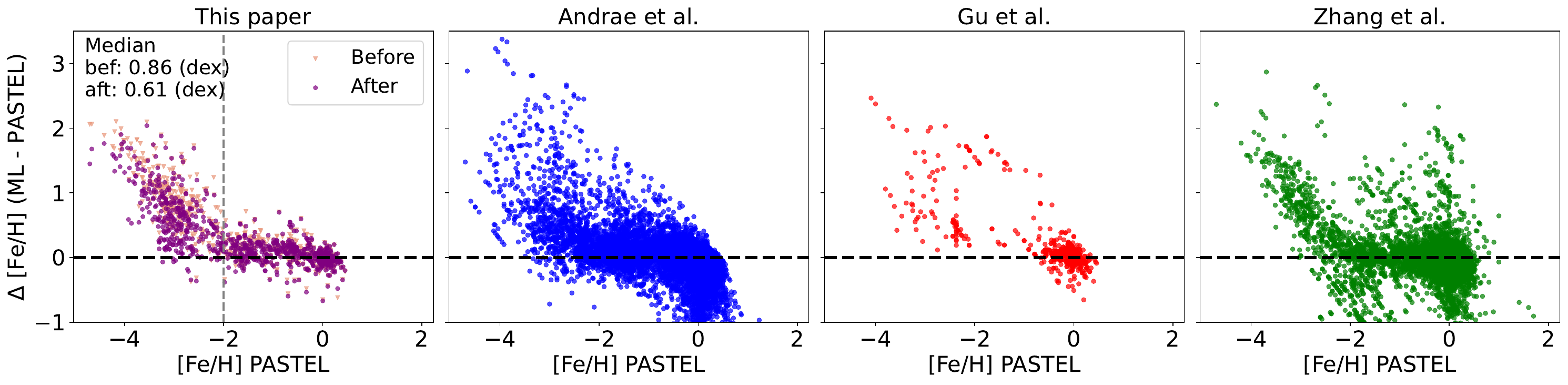}
   \includegraphics[width=\textwidth]{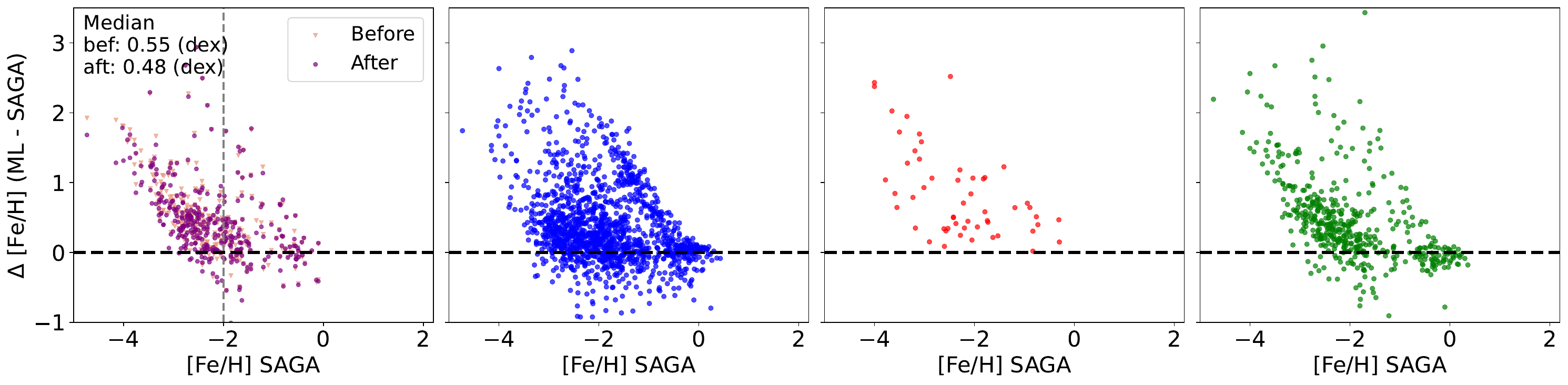}
      \caption{Comparison of the [Fe/H] predicted by this work and by three literature ML catalogs with the PASTEL (top row) and SAGA (bottom row) databases. Our results are shown in the leftmost plots, where the grey points show the results before, and the purple ones after, augmenting the reference SoS-Spectro catalog with very metal-poor stars. The median [Fe/H] differences for the before and after augmentation samples (computed only for [Fe/H] < –2) are indicated in the top-left corner of the first plot in each row. A similar comparison is presented for the following catalogs: \citet[][ blue, center-left panels]{andrae23}, \citet[][ red, center-right panels]{gu25}, and \citet[][ green, rightmost panels]{Zhang2023}. }
   \label{fig:metal_sum}
   \end{figure*}

If spectroscopic surveys and ML methods have trouble predicting [Fe/H] for metal-poor stars, in particular giants, this is even more pronounced in the case of very metal-poor stars, i.e., stars with [Fe/H]\,$\lesssim$\,--2\,dex. Fig.~\ref{fig:metal_sum} illustrates the problem for our catalog as well as the ones by \citet{Zhang2023}, \citet{andrae23}, and \citet{gu25}. The effect is observed both in the comparison with PASTEL and with the SAGA database \citep{suda08}. We tried to tackle the problem by augmenting the training sets for SDSS and SM with a few hundred very metal-poor stars selected from PASTEL (Sect.~\ref{sec:input}). The results before (grey) and after (purple) adding the stars are shown in the left panels of Fig.~\ref{fig:metal_sum}. As can be seen, their addition to the training sample did not solve the problem with the very metal-poor stars, but it did slightly improve our [Fe/H] estimates over the entire metallicity range. 

\subsection{White dwarfs} 

\begin{figure}[t]
   \centering
   \includegraphics[width=\columnwidth]{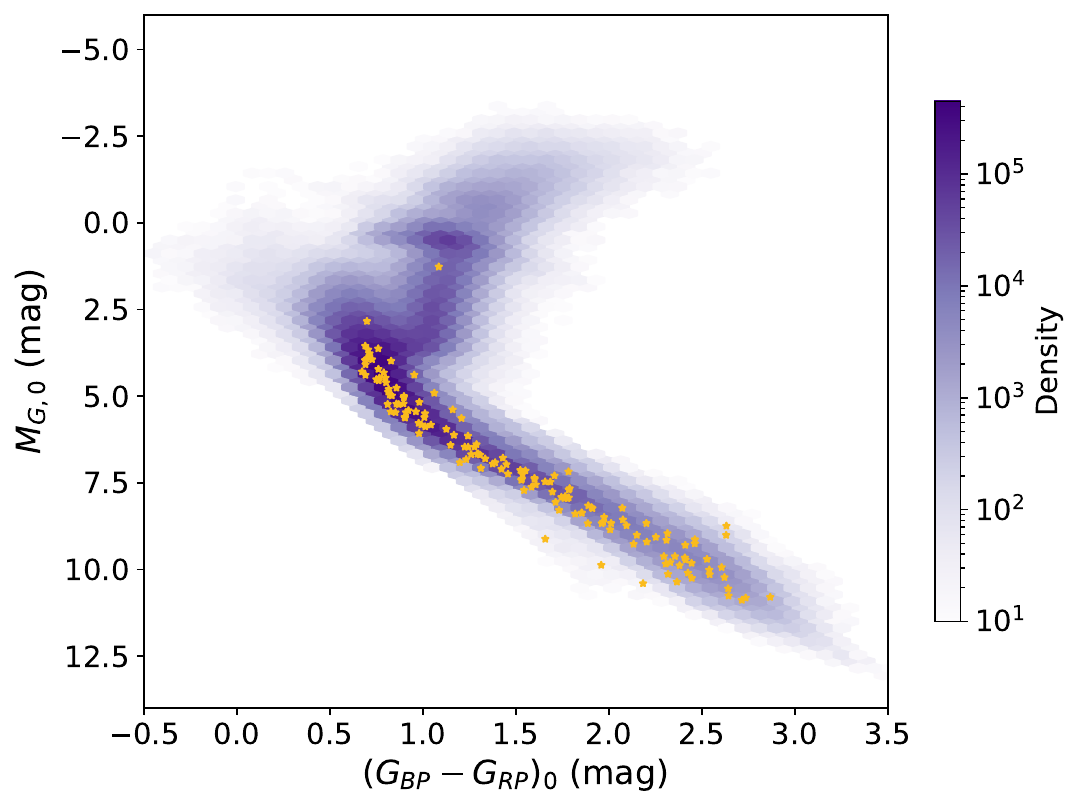}
      \caption{HR diagram showing the distribution of objects from our final catalog. $M_G$ is calculated using parallax. Yellow stars indicate 200 single UV-excess objects from \cite{Jackim2024} found in our catalog. }
   \label{fig:cmdfull2}
   \end{figure}

Our training sample does not include any white dwarf star (WD), but it is reasonable to expect that some WD might be present in the photometric catalogs and they could in principle have been wrongly parameterized. To verify that there are no WD in our final catalog, we cross-matched it with two published WD catalogs. The first is the \cite{Fusillo2021} catalog, which contains more than 1.2 million WDs and is based on {\em Gaia} DR3. A cross-match with this catalog yielded no common objects, suggesting that our catalog is indeed free from WDs. The second is the recently published single and binary WD catalog by \cite{Jackim2024}, which identifies single and binary WDs using GALEX UV color magnitude diagrams. The cross-match identified 40,276 objects in our final catalog, only 200 of which are classified as single WDs by \cite{Jackim2024}. However, when these objects are plotted on the HR diagram, they fall outside the WD locus (see Fig.~\ref{fig:cmdfull2}). Instead, they are distributed in regions associated with main-sequence stars or other stellar populations. This suggests that these objects are not single WD, but likely the non-WD components of binary systems, or alternatively stars that display UV excess because they are chromospherically active. Notably, none of these objects exhibit increased errors or abnormal properties, suggesting that their UV excess does not significantly impact their optical properties. 

\subsection{Gravity comparison with APOKASC}
The third APOKASC (APOKASC-3) catalog \citep{apokasc3} provides a comprehensive dataset for 15808 evolved stars, combining spectroscopic parameters from APOGEE with asteroseismic measurements from NASA's Kepler mission \citep{kepler}. Of these, detailed parameters such as stellar evolutionary state, surface gravity, mass, radius, and age are available for 12418 stars, calibrated using {\em Gaia} luminosities and APOGEE spectroscopic effective temperatures. The catalog features precise asteroseismic measurements with median fractional uncertainties of 0.6\% in \(\nu_{\mathrm{max}}\) and \(\Delta\nu\), and 1.8\% in radius, making it a reliable resource for validating derived stellar parameters. Given its robust calibration and high data quality, APOKASC-3 is well-suited for testing log\,$g$ values, providing a consistent and independent benchmark for stellar characterization studies.

\begin{figure}[t]
   \centering   \includegraphics[width=\columnwidth]{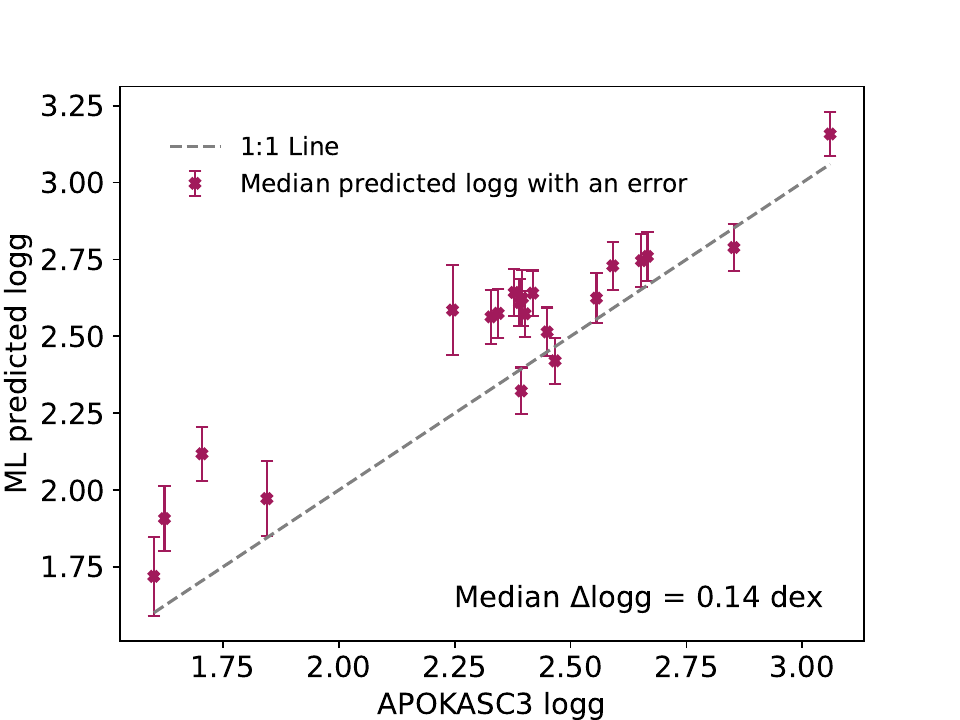}
      \caption{Comparison of surface gravity log\,$g$ values by APOKASC-3 with those predicted using machine learning in this study. The error bars represent the uncertainties in our predictions.}
   \label{fig:apokasc_logg}
   \end{figure}

Fig.~\ref{fig:apokasc_logg} shows the comparison between the surface gravity log\,$g$ values estimated in APOKASC-3 and those predicted in our work. The error bars on our predictions represent the associated uncertainties. As can be seen, a systematic offset is present, albeit the few stars in common, with a median difference of 0.14~dex and with the errorbars barely touching the 1:1 line. The effect is visible especially for giants with log\,$g$\,$\lesssim$\,2.5\,dex. This difference is consistent with the offset observed between the SoS-Spectro dataset (used as a reference for ML training) and high-quality log\,$g$ measurements from the PASTEL catalog (see Fig.\ref{fig:pastel1}). Such discrepancies likely results from general inconsistencies in the determination of log\,$g$ for giant stars across large spectroscopic surveys, which are impacted also by NLTE effects, especially at low metallicity. Our recalibration of the SoS-Spectro on PASTEL was not sufficient to erase this trend with log\,$g$, unlike what happened for [Fe/H] in the case of globular clusters. This remains one important challenge for the future SoS data releases and for spectroscopic surveys and ML methods in general.

\subsection{Validation of cool star temperatures using spectral types}

   \begin{figure}[t]
   \centering   \includegraphics[width=\columnwidth]{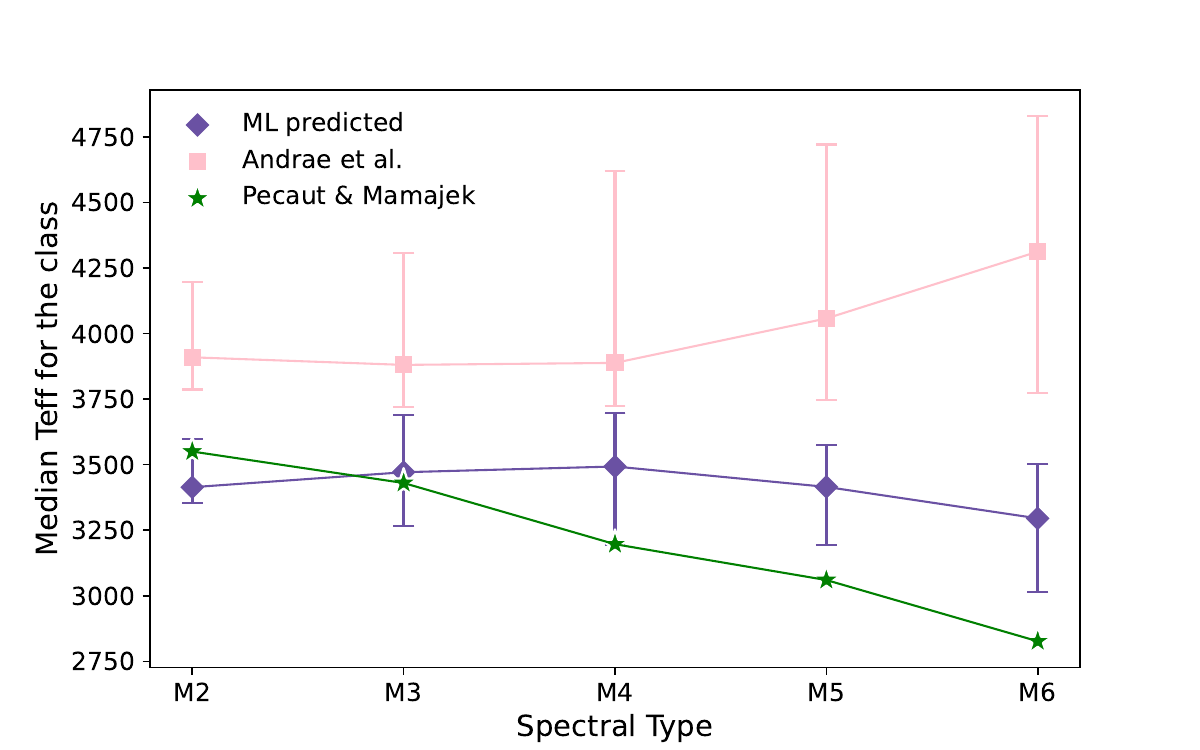}
      \caption{Calculated median effective temperatures and their range (whiskers) for each spectral type from Andrae et al. (pink squares) and as predicted in this paper (purple diamonds). The values provided by Pecaut \& Mamajek are shown as green stars.
}
   \label{fig:andrae_expl}
   \end{figure}

Among the most striking features in the comparisons presented in Sect.~\ref{sec:val} is the difference in T$_{\rm{eff}}$ with \cite{andrae23}, particularly for the cool stars in our catalog. To explore this discrepancy, we selected stars with ML-predicted $T_{\rm eff} < 3400$~K and cross-matched them with the Simbad database\footnote{\url{https://simbad.cds.unistra.fr/simbad/}}. Among these, we identified approximately 700 stars with spectral types ranging from M2 to M6, the majority being classified as M4V. We grouped the spectral types into rounded categories and calculated the median effective temperature for each group using both our ML predictions and the estimates of \cite{andrae23}.

Fig.~\ref{fig:andrae_expl} presents these medians, along with the range of temperature estimates shown as whiskers. For reference, we also include the T$_{\rm{eff}}$ for each spectral type as provided by \cite{PecautMamajek2013}. It is important to note that these values are strictly valid for main-sequence stars, while our sample may contain a few giants and several pre-main sequence stars. Nonetheless, they offer a reasonable benchmark for T$_{\rm{eff}}$ estimates based on spectral type. In this regard, our results align more closely with the known spectral classifications.

\section{Cluster by cluster [Fe/H] comparisons}
\label{sec:per_cluster}

In this section, we present a cluster-by-cluster comparison of the [Fe/H] estimates derived in this work with those obtained from external ML-based studies. We adopt the same cluster membership selections and literature metallicity values as in Sect.~\ref{sec:val}. 

\subsection{Globular clusters}

\begin{figure*}[t]
   \centering
   \includegraphics[width=\textwidth]{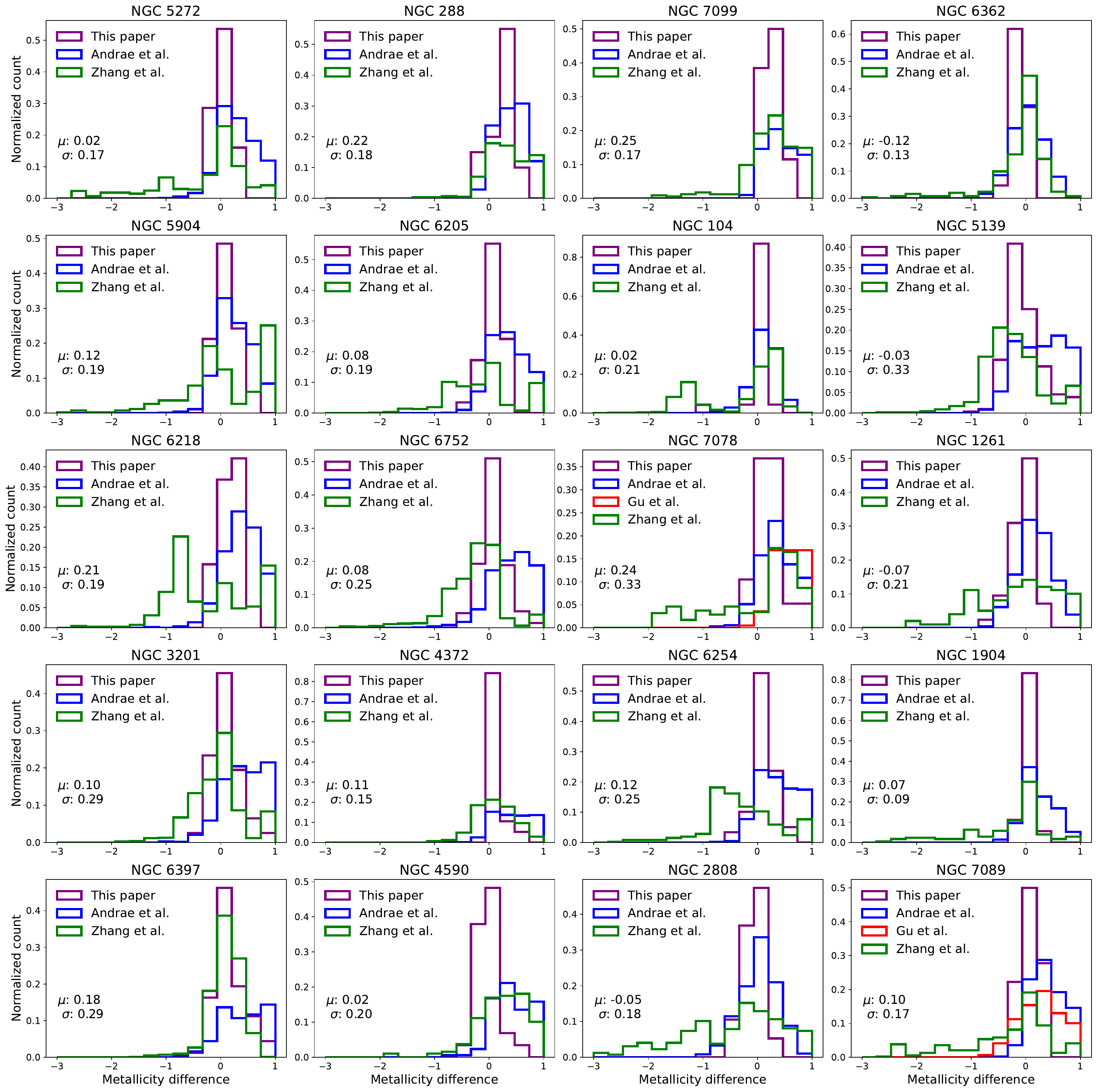}
      \caption{The [Fe/H] differences between predictions from various ML methods and \citet{Harris1996} reference values for globular clusters. Each panel represents a cluster, with histograms of [Fe/H] differences for the four ML methods explored in Sect.~\ref{sec:val}: this paper, \citet{andrae23}, \citet{Zhang2023}, and \citet{gu25}. The mean and standard deviation of the differences are annotated only for our method.}
   \label{fig:glob_cluster_all}
   \end{figure*}

Under these criteria, 20 globular clusters are retained, although only two (NGC 7078 and NGC 7079) have sufficient data from \citet{gu25} to be included. Figure~\ref{fig:glob_cluster_all} shows, for each group, the distribution of differences between the estimated [Fe/H] values and the metallicities in the \citet{Harris1996} catalog. Each subplot presents histograms of the residuals for each data source, allowing a direct visual comparison across ML methods. The histograms are normalized by the number of objects in each dataset to account for differences in sample size. As can be seen, our method generally produces narrower and more centered residual distributions, indicating improved consistency with reference metallicities.

Nevertheless, some clusters exhibit notable deviations from Harris, which merit further discussion. NGC\,288 shows systematically higher metallicities across all data sources, consistent with the results of \citet{2020MNRAS.492.1641M}, who report a 0.136 dex offset from Harris. For NGC\,7078 (M\,15), both the mean and scatter are elevated, which may reflect the large discrepancies found between LTE and NLTE abundance estimates in this cluster \citep{2019A&A...628A..54K}. In the case of NGC\,6218, \cite{Harris1996} reports [Fe/H] = –1.37, while APOGEE-based studies find systematically higher values (–1.26 to –1.27; \citealt{2020MNRAS.493.3363H, 2024MNRAS.528.1393S}), suggesting possible systematic offsets in either APOGEE or the \cite{Harris1996} compilation. It is worth mentioning that our results tend to align more closely with those reported by \citet{Schiavon2024} than with the \citet{Harris1996} catalog. This agreement may stem from the fact that recent catalogue of \citet{Schiavon2024} make use of APOGEE-based measurements, and our estimates are also based on spectroscopic surveys. For NGC\,3201, although our mean metallicity agrees with both Harris and \citet{Schiavon2024}, the large difference between their reference values (–1.59 and –1.39, respectively) may explain the broader residual distribution we observe. Similarly, NGC\,5139 ($\omega$ Cen) shows a higher dispersion, although this is expected due to its complex internal metallicity distribution.

A few clusters show sligthly larger spreads in our SoS-ML catalog. NGC\,6254 exhibits a larger scatter in our results compared to other surveys. A similar trend was noted by \citet{2025A&A...693A.155P}, who reported a dispersion of 0.07 dex, slightly lower than what we observe. However, their estimate assumes that the ASPCAP uncertainties are reliable. This assumption may not hold, as previous analyses have shown that APOGEE's reported uncertainties can be significantly underestimated, with the three-cornered hat method revealing much larger true errors \citep{tsantaki22}.  NGC\,7099 and NGC\,6752 both display slightly elevated or broadened residuals across data sources without any clear explanation in the literature. 

\subsection{Open clusters}

Following the criteria described above, we used the members list by \citet{HuntReffert2024} and compared our results with the homogeneous mean cluster metallicities compiled by \citet{Netopil2016}, based on high-resolution spectroscopy. We used 16 open clusters with more than 20 bona fide members. As reported in Sect.~\ref{clustervalid}, all the ML methods show an average offset with the reference mean [Fe/H] for the selected open clusters. The same offset remained when comparing against different catalogs, such as \citet{kharchenko16} or \citet{dias21}.

\begin{figure*}[t]
   \centering
   \includegraphics[width=\textwidth]{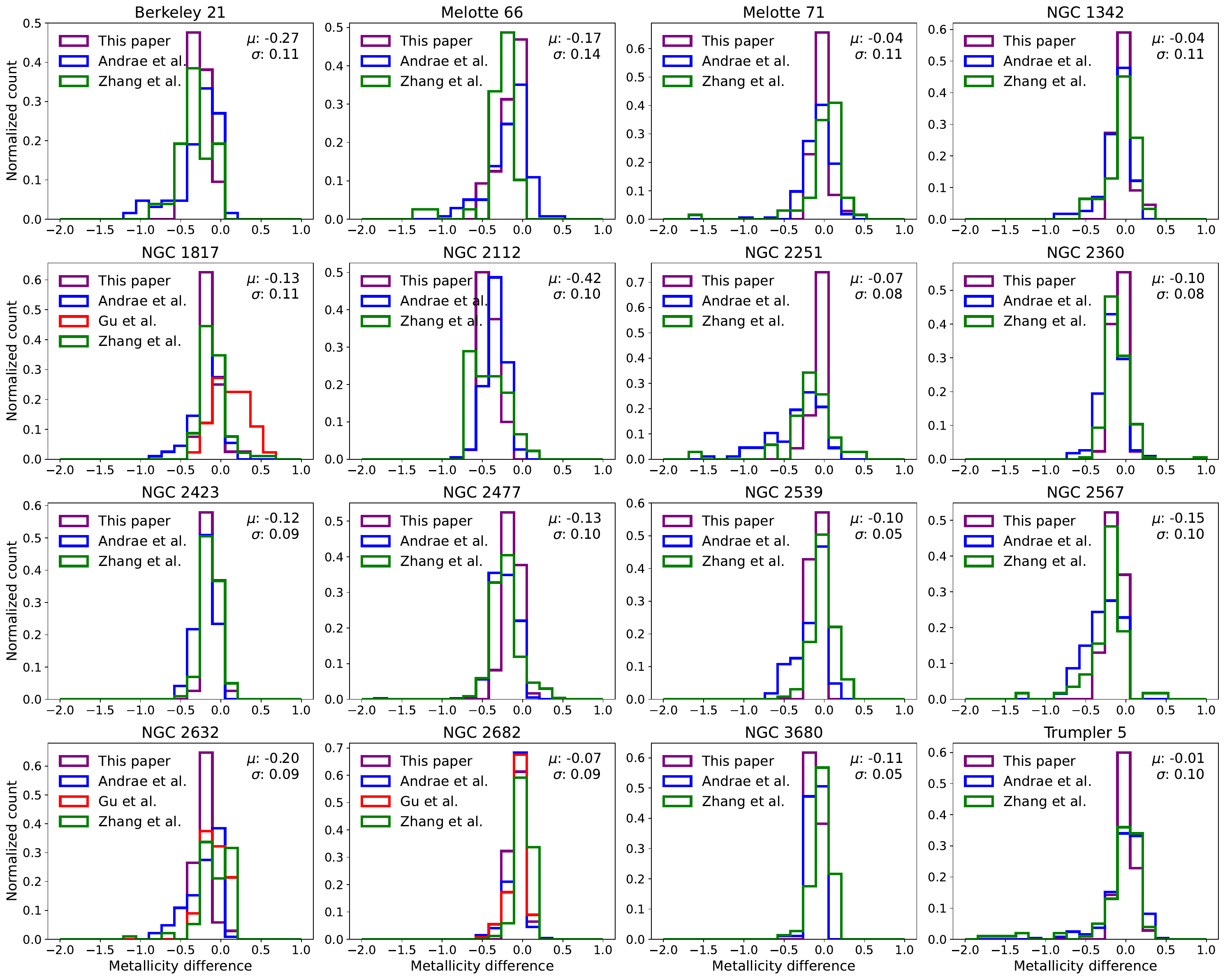}
      \caption{The same as in Fig.\ref{fig:glob_cluster_all}, but for open clusters.}
   \label{fig:open_cluster_all}
   \end{figure*}

The cluster-by-cluster comparisons are shown in Fig.~\ref{fig:open_cluster_all}. They generally confirm the agreement among the ML methods, with the only exception of NGC\,1817, where the results by \citet{gu25} appear to have a smaller bias and a larger spread. We also note a general tendency of our predictions to have a smaller spread and fewer outliers in the distributions of each cluster, similarly to what observed for globular clusters, but in a less pronounced way. This confirms that we could improve slightly on the precision of the predicted [Fe/H] values in the range of Solar metallicity, but not on the bias, compared to other methods.  

\end{appendix}

\end{document}